\documentclass[aps,prl,reprint,superscriptaddress]{revtex4-2}

\usepackage{graphicx}% Include figure files
\usepackage{dcolumn}% Align table columns on decimal point
\usepackage{bm}% bold math

\usepackage{amsmath}
\usepackage{amsfonts}
\usepackage{amssymb}
\usepackage{mathtools}

\usepackage{xcolor}
\usepackage{array}

\usepackage{placeins}

\newcommand{\ket}[1]{\left\vert #1 \right\rangle}
\newcommand{\Fout}{F_{\text{out}}}

\usepackage{hyperref}% add hypertext capabilities
\begin{document}

\title[Spectral Engineering of Cavity-Protected Polaritons in an Atomic Ensemble]{Spectral Engineering of Cavity-Protected Polaritons \\ in an Atomic Ensemble with Controlled Disorder}

\author{Mohamed Baghdad}
\thanks{These authors contributed equally to this work.}
\affiliation{Laboratoire Kastler Brossel, ENS-Universit\'{e} PSL, CNRS, Sorbonne Universit\'{e}, Coll\`{e}ge de France, 24 rue Lhomond, Paris, 75005,  France}

\author{Pierre-Antoine Bourdel}
\thanks{These authors contributed equally to this work.}
\affiliation{Laboratoire Kastler Brossel, ENS-Universit\'{e} PSL, CNRS, Sorbonne Universit\'{e}, Coll\`{e}ge de France, 24 rue Lhomond, Paris, 75005,  France}

\author{Sylvain Schwartz}
\thanks{These authors contributed equally to this work.}
\affiliation{Laboratoire Kastler Brossel, ENS-Universit\'{e} PSL, CNRS, Sorbonne Universit\'{e}, Coll\`{e}ge de France, 24 rue Lhomond, Paris, 75005,  France}
\affiliation{Present address: DPHY, ONERA, Universit\'{e} Paris-Saclay, 91123 Palaiseau, France.}

\author{Francesco Ferri}
\affiliation{Laboratoire Kastler Brossel, ENS-Universit\'{e} PSL, CNRS, Sorbonne Universit\'{e}, Coll\`{e}ge de France, 24 rue Lhomond, Paris, 75005,  France}

\author{Jakob Reichel}
\affiliation{Laboratoire Kastler Brossel, ENS-Universit\'{e} PSL, CNRS, Sorbonne Universit\'{e}, Coll\`{e}ge de France, 24 rue Lhomond, Paris, 75005,  France}

\author{Romain Long}
\email{long@lkb.ens.fr} 
\affiliation{Laboratoire Kastler Brossel, ENS-Universit\'{e} PSL, CNRS, Sorbonne Universit\'{e}, Coll\`{e}ge de France, 24 rue Lhomond, Paris, 75005,  France}

\date{\today}

%%==================================%%
%% abstract %%
%%==================================%%

\begin{abstract}
The paradigm of $N$ quantum emitters coupled to a single cavity mode appears in many situations ranging from quantum technologies to polaritonic chemistry. The ideal case of identical emitters is elegantly modeled in terms of symmetric states, and understood in terms of polaritons. In the practically relevant case of an inhomogeneous frequency distribution, this simple picture breaks down and new and surprising features appear. Here we leverage the high degree of control in a strongly coupled cold atom system, where for the first time the ratio between coupling strength and frequency inhomogeneities can be tuned. We directly observe the transition from a disordered regime to a polaritonic one with only two resonances. The latter are much narrower than the frequency distribution, as predicted in the context of ``cavity protection''. We find that the concentration of the photonic weight of the coupled light-matter states is a key parameter for this transition, and demonstrate that a simple  parameter based on statistics of transmission count spectra provides a robust experimental proxy for this theoretical quantity. Moreover, we realize a dynamically modulated  Tavis-Cumming model to produce a comb of narrow polariton resonances protected from the disorder, with potential applications to quantum networks.
\end{abstract}




\maketitle

While the foundations of cavity quantum electrodynamics (CQED) have been laid with single-emitter systems \cite{Kimble1998, Haroche2006}, the extension to many emitters coupled to the same cavity mode holds interest for many different communities \cite{ Solnyshkov2021, Blais2020b, Chang2018a, Ebbesen2016}. In solid-state systems, with emitters such as rare earth ions, color centers in diamond, or semiconductor quantum wells, the $N$-emitter situation arises naturally and is exploited to enhance the light-matter coupling beneficial for quantum network implementations \cite{Bussieres2013, Heshami2016a, Zhong2017b}. In atomic systems, the coupling between emitters via the cavity mode is a powerful approach for generating different forms of many-particle entanglement \cite{Haas2014, Mcconnell2015, Welte2017a} including spin-squeezed states for quantum metrology \cite{Leroux2010, Schleier-smith2010, Ma2011, Hosten2016, Pezze2018}. When combined with a high degree of control over the emitter qubits, this is the basis for an emerging form of quantum simulations of long-range coupled spin systems \cite{Baumann2010a, Chang2018a, Kroeze2018, Muniz2020, Periwal2021a}. All these experiments operate in the regime of low excitation, with no more than one photon present in the cavity, and are performed in the regime of strong collective coupling, or even strong single-emitter coupling. 

Central to such systems is the notion of a polariton \cite{Basov2020}, the hybrid light-matter state composed of an intracavity photon and the ensemble of emitters sharing a single excitation. In homogeneous systems where all emitters have the same resonance frequency $\omega_0$, the energy spectrum of the coupled system is the well-known avoided crossing with only two resonances. Although many more, less symmetric atomic states exist, they do not participate in the system's transmission or reflection spectra because they are ``dark'': their collective coupling to the photonic mode vanishes. 
However, as soon as there is frequency inhomogeneity in the emitter ensemble, this destructive interference is no longer complete, and the formerly dark states now couple to the light field, becoming ``gray''. While frequency inhomogeneity is easily included in the foundational Tavis-Cummings Hamiltonian \cite{Tavis1968}, it leads to rich and often surprising physics, as shown by many recent advances driven by the interplay of theory and experiment. One spectacular example in the field of materials science started with the discovery that coupling an organic semiconductor to a cavity can profoundly change its material properties, and in particular, enhance its conductivity through the hybridization of electronic transitions \cite{Orgiu2015}. Recent theoretical advances point out the importance of the ``gray'' states in the localization and transport properties of these systems \cite{schachenmayer2015, Gonzalez-ballestero2016,   Botzung2020a, Chavez2021}.

In this context, atomic CQED systems with their high degree of control appear as a promising tool for quantum simulation \cite{Mivehvar2021} of such CQED-enhanced materials. Another prominent example is the phenomenon of ``cavity protection'': a system with strong inhomogenous broadening can still support narrow resonances (much narrower than the width of the broadening) under certain conditions \cite{Houdre1996a, Kurucz2011a, Diniz2011b}. Cavity protection has been experimentally demonstrated recently in the context of solid-state CQED  \cite{Putz2014a, Breeze2017, Zhong2017b}, where inhomogenous broadening is dictated by material properties and the narrow resonances warrant hope for quantum memories with long coherence times despite fast dephasing of the uncoupled emitters. Clearly, the features of such a system depend on the relative strength of the broadening. However, this is usually not adjustable in solid-state systems, precluding a direct observation of the transition to the protected regime, from which a better understanding of the physics would emerge.

In this article, we present results of an atomic CQED experiment with tunable atomic frequency distribution. We experimentally show the transition from a ``protected'' regime where only two polaritonic resonances are present despite an inhomogeneous frequency distribution that is much larger than the polariton resonances, to an ``unprotected'' one where the amplitude of the polaritons decreases and many additional, randomly distributed resonances appear. Our analysis shows that the number of photon counts outside the two narrow polariton peaks acts as a robust experimental proxy that closely traces the calculated photonic weight of the gray states.
In addition, as our system operates in the strong coupling regime at the single atom level, we are able to study this transition for a few tens of atoms, several orders of magnitude lower than previous experiments, highlighting the contribution of a finite number of dark states. Finally, we produce polaritons featuring a comb-like frequency spectrum by modulating the light-shifted frequency of the atoms on a very fast time scale in the protected regime. This allows us to shape the frequency spectrum of the system, with possible applications to quantum memories and quantum communications \cite{Lukin2020a, Craiciu2021, Mishra2021a, White2022}.

\begin{figure}
\centering
\includegraphics[width=\columnwidth]{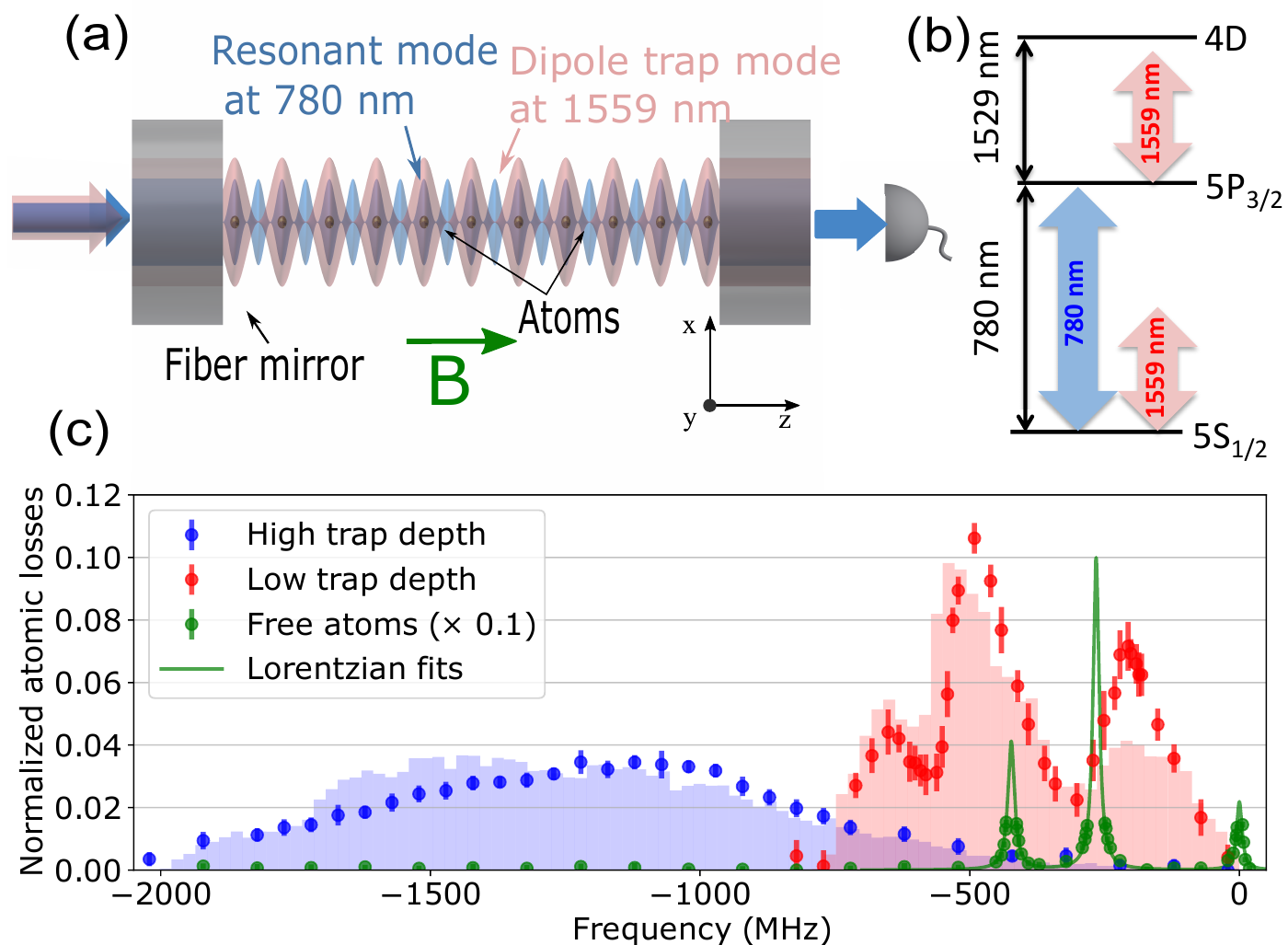}
\caption{\textbf{Experimental setup of the cold atom cavity interface with tunable inhomogeneous frequency distribution.}  (a) Sketch of the setup. Atoms are trapped in a one-dimensional optical lattice (red) which is commensurate with the main cavity mode (blue). The latter is on resonance with the $\ket{g}\leftrightarrow\ket{e}$ atomic transition.  (b) Simplified level diagram illustrating the proximity of the $1559\,$nm trapping light with the $5P_{3/2} \leftrightarrow 4D$ line. (c) Atomic frequency distribution measured by loss spectroscopy of the trapped atoms for various trap depths (Green: no trap. Red: $U=310\,\mu$K. Blue: $U = 1400\,\mu$K). The horizontal axis corresponds to the frequency of the transverse probe beam inducing the losses, referenced to the bare $\ket{5S_{1/2},F=2,m_F=2} \leftrightarrow \ket{5P_{3/2},F'=3,m_{F'}=3}$ transition. Circles: experimentally measured losses (error bars correspond to $1\sigma$  spread of the data), normalized to the maximum value of the $F'=2$ peak of the free atoms. Histograms: result of Monte Carlo simulations of the frequency distributions (see main text and Supplementary Information).}

\label{fig:Figure1}
\end{figure}

\section{Cold Atom Cavity Interface with Tunable Frequency Inhomogeneity}

In our experiment, a cold ensemble of Rubidium atoms is trapped inside a high-finesse Fabry-Perot microcavity \cite{Hunger2010} with a length of $140\,\mu$m. The number of atoms loaded into the cavity can be varied between a few tens and a few thousands (see Methods). Unless otherwise noted,  atoms are initially prepared in $\ket{g}=\ket{5S_{1/2},F=2, m_F=2}$. The probe beam exciting the cavity mode on the D2 line of $^{87}$Rb at 780$\,$nm is circularly polarized, coupling $\ket{g}$ to the excited state  $\ket{e}=\ket{5P_{3/2},F'=3, m_F'=3}$. The choice of circular polarization is possible thanks to the negligible birefringence of our cavity \cite{Garcia2018}. To realize a homogeneous and maximal coupling of all the atoms, they are trapped at the antinodes of the cavity mode. This is achieved by an intracavity standing wave at $\omega_t=2\pi c_0/1559\,$nm where $c_0$ is the speed of light, which is commensurate with the atomic line taking into account the Gouy phase shift and the phase shift at reflection on the mirrors \cite{Garcia2020a, Ferri2020a}. The dipole light is linearly polarized and the trap depth of the intracavity lattice can be adjusted between 300$\,\mu$K and 1400$\,\mu$K with atomic temperatures ranging from 50$\,\mu$K  to 160$\,\mu$K. The maximum single-atom coupling is $g_0/2\pi\simeq 76\,$MHz; thermal motion leads to a temperature-dependent average single-atom coupling $g/2\pi$ between 55 and 68$\,$MHz (see Supplementary Information). This is larger than the cavity decay rate (HWHM) $\kappa/2\pi \simeq 15\,$MHz and the spontaneous emission rate $\gamma/2\pi \simeq 3\,$MHz, placing the system in the strong coupling regime of CQED.

Due to a nearby resonance (the $5P_{3/2} \leftrightarrow 4D$ transition at 1529$\,$nm), the trapping light induces a light shift of the excited state of the D2 line which is $\simeq 50$ times larger than the ground state light shift \cite{Brantut2008}. Thus, it creates an adjustable shift and broadening of the frequency distribution of the $\ket{g} \leftrightarrow \ket{e}$ transition, the latter depending also on the finite temperature of the trapped atoms (see Methods). We characterize the frequency distribution experimentally by illuminating the trapped atoms with a transverse beam and measuring the relative atomic losses as a function of the beam frequency for various trap depths, as shown on Fig.~\ref{fig:Figure1} (c). It shows the broadening (up to 1\,GHz) and shift due to  the  the 1559$\,$nm light power. A Monte Carlo simulation assuming an initial atomic population uniformly distributed between all the Zeeman sublevels, but no subsequent redistribution or depumping (see Methods), is in qualitative agreement with the experimental loss curves.

\section{Cavity Protection Effect}

To investigate the coherence properties of this inhomogeneous system, we consider the transmission spectrum of the cavity in the low excitation limit, which is relevant for our experimental configuration. 
We start by describing the homogeneous case. For a cavity mode resonant with the atomic line in the strong coupling regime, the spectrum exhibits the well-known vacuum Rabi splitting featuring two peaks $2 g\sqrt{N}$ apart in frequency, where $N$ is the atom number, and their width is given by  $(\kappa+\gamma)/2$ \cite{Haroche2006}. They correspond to two polaritons of the form:
\begin{equation}
\vert P_\pm \rangle=\frac{1}{\sqrt{2}} \left(\vert 1, G\rangle \pm \vert 0, W \rangle \right),
\end{equation}
where $\ket{1, G}$  is the state with one photon in the cavity mode and all atoms in the ground state and $\ket{0, W}$ is a state with zero photons in the cavity mode and one excitation symmetrically shared between the atoms. In addition, $N-1$ degenerate dark states are not coupled to the cavity light field. This is reflected in the photonic weight (PW), which is defined by the overlap of the eigenstates of the coupled system $\vert\psi_\alpha\rangle$ with $\ket{1, G}$, 
$PW= \vert \langle G,1 \vert \psi_\alpha \rangle  \vert^{2}$ \cite{Dubail2022}, and is zero for all eigenstates except $\ket{P_\pm}$.

In our experiment, two types of inhomogeneities are present. First, each atom $k$ has a different value of the coupling $g_k$ to the cavity. This does not change the previous result, except that the splitting is now given by $2 \Omega$, where $\Omega=\sqrt{\sum_{k=1}^N g_k^2}$ is the collective coupling and $\ket{W} = \Sigma^{\dagger} \ket{G}$ with $\Sigma^{\dagger} = \sum_{k=1}^N \frac{g_k}{\Omega} \sigma_{k}^{+}$, $ \sigma_{k}^{+}$ being the raising operator for atom $k$. 
Secondly, each atom has a different light-shifted atomic frequency $\omega_k$ depending on its motional state in the trapping potential and this leads to more profound modifications. In the strong coupling limit, two hybrid light-matter collective modes (denoted simply polaritons in the following) still exist and are split by $\simeq 2 \Omega$. However, their coherence is limited as they are coupled to the $N-1$ other eigenstates, that we denote \textit{dark states} in the following as they carry zero photonic weight for infinite coupling. This coupling can be strongly reduced if there is an energy gap between the polaritons and the dark states. This is achieved when the energy gap, given by $\Omega$, is much larger than the atomic frequency distribution width $\Delta \omega$ \cite{Houdre1996a, Kurucz2011a, Diniz2011b}.

\begin{figure}[t]
\centering
\includegraphics[width=\columnwidth]{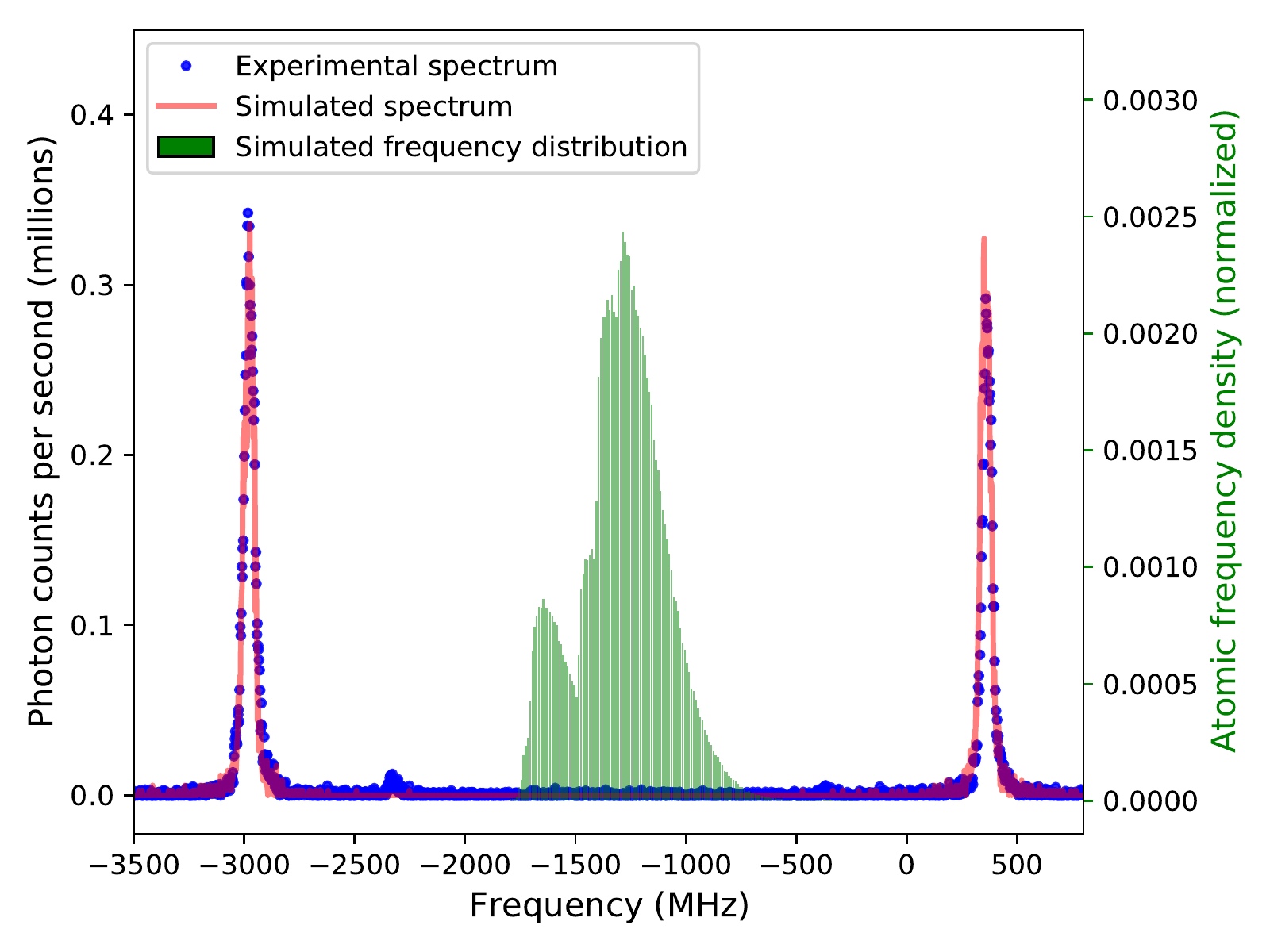}
\caption{ \textbf{Cavity protection effect}. Blue dots represent the experimental on-resonance transmission spectrum in the regime where the collective coupling is much larger than the atomic frequency distribution. The two polariton modes appear as two narrow transmission peaks, in good agreement with the numerical simulation (red line). Green bars show the simulated atomic frequency distribution, which is much broader than the transmission peaks.}
\label{fig:Figure2}
\end{figure}

For our specific system, we compute the spectral distribution of the coupling of the atomic ensemble with the cavity field: $\rho(\omega)=\sum_{k=1}^N \sum_{j=1}^{16} g_{k,j}^{2} \delta(\omega-\omega_{k,j})$, where we sum over all 16 excited sublevels of the 5P$_{3/2}$ manifold (see Supplementary). The distribution is shown in Fig.~\ref{fig:Figure2} for a trap depth of $\approx 1400\,\mu$K. It has an asymmetrical shape and features a bounded support $1700\,$MHz wide, which corresponds to the highest possible value of the light shift for this trap depth. The frequency $\omega_c$ of the cavity mode is tuned to resonance with the mean value of the frequency distribution. The experimental transmission spectrum of the cavity averaged over $\approx 200$ spectra is also represented in Fig. \ref{fig:Figure2}. Despite the broad atomic frequency distribution, it features only two peaks corresponding to the polaritons. To first order in $\Delta \omega/ \Omega$ with $\Delta \omega$ the width of the frequency distribution, they are split by twice the collective coupling $\Omega/2 \pi \simeq 1670\,$MHz, which corresponds to $N\approx 770$ atoms trapped in the cavity mode. In this regime, we fully benefit from the ``cavity protection'' effect: the two polaritons lying outside the distribution support, they are decoupled from the dark states and the coherence of the system is preserved. 

The fitted width of the peak $\delta \omega/2 \pi = 26 \pm 2\,$MHz is much lower than the width $\Delta \omega/2 \pi \approx 150\pm10\,$MHz of the frequency distribution (see Methods). The ratio $\frac{\Delta \omega /2}{\delta \omega}$ can be used to define a figure of merit of the cavity protection. It allows the comparison between the measured width $\delta \omega$ of the polaritonic peaks and the width $\Delta \omega/2$ that would be obtained for a Lorentzian probability distribution inhibiting the protection effect \cite{Houdre1996a, Kurucz2011a, Diniz2011b}. This ratio is about 3 for our experiment, showing that despite strong inhomogeneities, the coherence of the polaritons can be preserved as long as $\Omega \gg \Delta \omega $ for our frequency distribution.  We note that the measured widths are larger than the homogeneous limit of $(\kappa+\gamma)/2 \simeq 9\,$MHz, mainly due to shot-to-shot fluctuations of the atom number, which broadens the averaged peaks by inducing a jittering of the collective coupling. The widths are indeed the same for the different trap depths used in the experiment within the error margins. As can be seen on Fig.\ref{fig:Figure2}, the simulated spectra including atom number fluctuations are in excellent agreement with the experimental data.

\section{Transition from the unprotected to the protected regime }

\begin{figure}[t]
\centering
\includegraphics[width=\columnwidth]{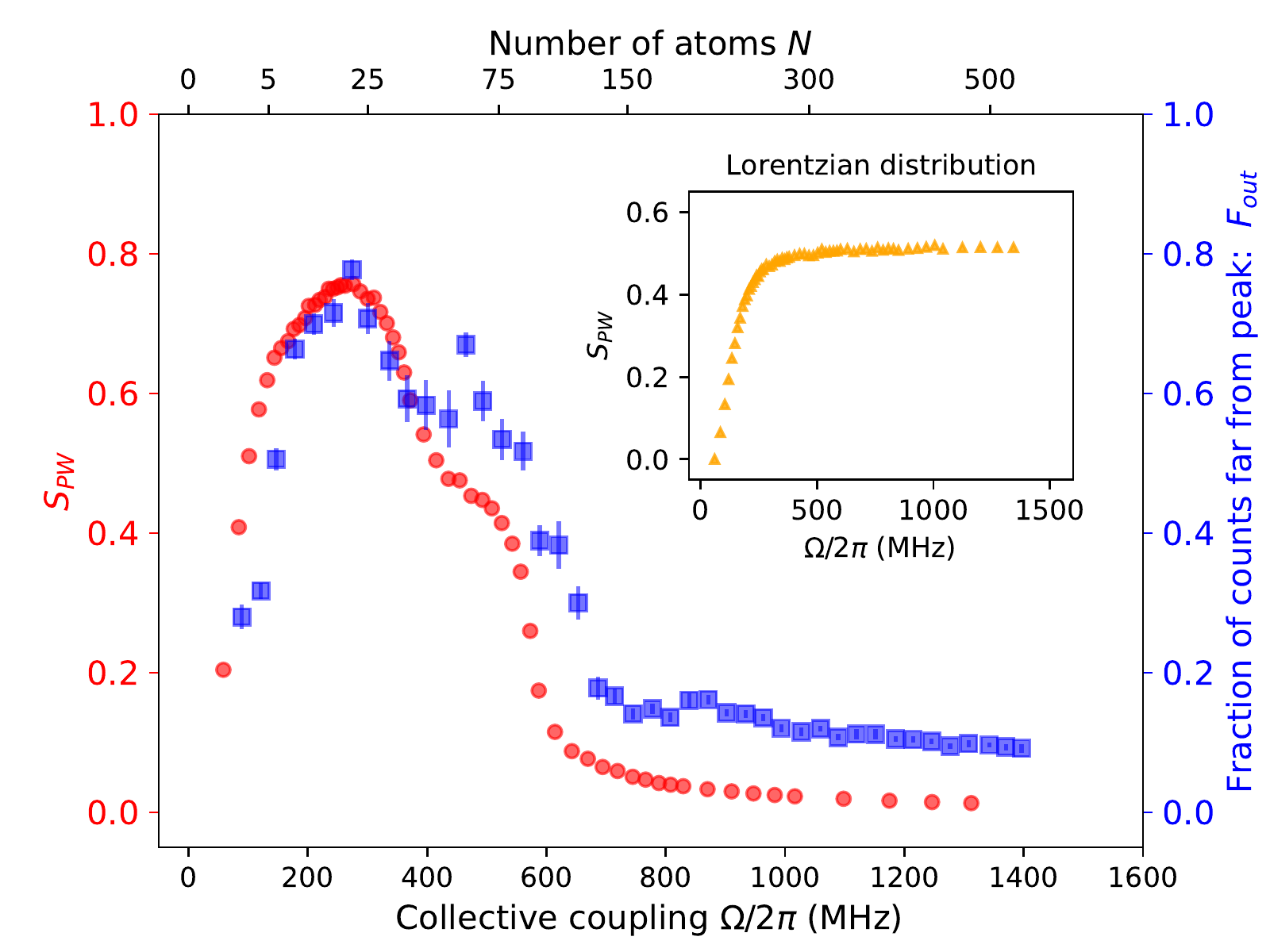}
\caption{ {\textbf{Transition from the unprotected to the protected regime}. For a given atomic frequency distribution, we  increase the number of atoms and thus the collective coupling.} Blue squares: measured $\Fout$. Beyond a collective coupling $\Omega_{th}/(2 \pi) \simeq 300\,$MHz, the fraction of photon counts outside the two polariton peaks rapidly drops. $\Fout$ is closely traced by the simulated photonic weight $S_{PW}$ (red circles): confirming that the off-peak transmission corresponds to gray states whose photonic weight drops as faster coupling wins over disorder-induced dephasing.  For comparison, the inset shows $S_{PW}$ in the case of a Lorentzian resonance frequency distribution. Since no cavity protection occurs in this case, $S_{PW}$ remains high at large collective coupling. 
}
\label{fig:Figure3}
\end{figure}

\begin{figure*}[t]
\centering
\includegraphics[width= \textwidth]{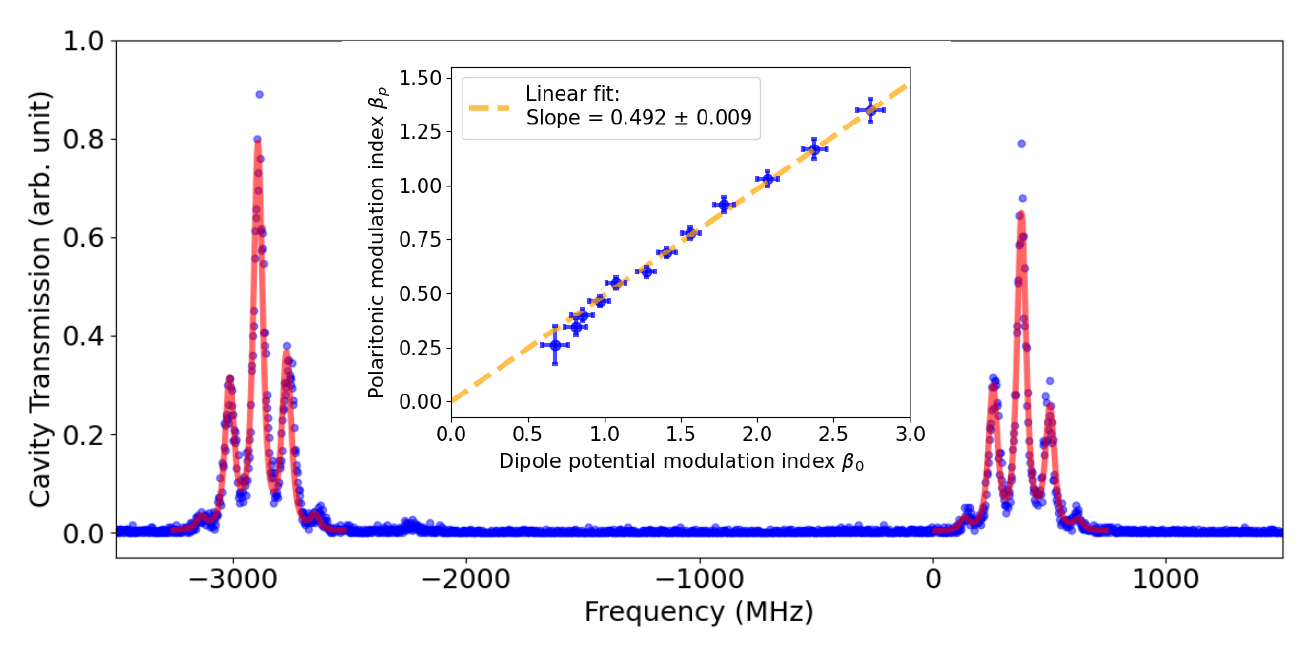}
\caption{\textbf{Modulation of the polariton eigenfrequencies}. Blue points: experimental data with $\Omega/2 \pi = 1630\pm  20\,$MHz, $\omega_m /2\pi =  120\,$MHz, $\beta_o = 2.17\pm0.04$. 
Red line: simulated spectrum obtained by numerically integrating the master equation. Inset: Modulation transfer function. Blue points: experimental data of the modulation index of the polaritons $\beta_p$ for different values of the modulation index $\beta_o$ of the atomic frequency. Red dashed line: linear fit of data points gives a slope of $0.492 \pm 0.009$, compatible with the theoretical slope of 0.5.}
\label{fig:Figure4}
\end{figure*}

Exploiting the high degree of control of our system, we explore the transition from the unprotected to the protected regime. To exhibit this transition, we acquire cavity transmission spectra as a function of the number $N$ of trapped atoms in the cavity mode, for $10\lesssim N \lesssim 550$. As we are dealing with a mesoscopic number of atoms, we can compute the photonic weight of the different eigenstates of the coupled system for the trap depth and atom temperature of the experiment. The cavity protection effect can directly be assessed by measuring the distribution of the PWs over the different eigenstates. For a system with inhomongeneity of the emitter frequencies, the $PW$ is distributed over more than two eigenstates. This leads us to introduce the sum $S_{PW}$ of the $PW$ of all the eigenstates except the two largest ones, to quantify the spreading of the $PW$ and thus the coherence of the coupled system: $S_{PW}$ is zero for the homogeneous case and tends to 1 when the $PW$ is distributed over an infinite number of dark states.

Experimentally,  as we stay in the low-excitation limit, the small number of collected photons per spectrum prevents the direct measurement of the PW of the eigenstates. Instead, we extract from each spectrum the fraction $\Fout$ of photon counts that are outside of two narrow frequency windows corresponding to the polariton peaks (see Methods). The width  $\Delta f$ of these windows is chosen such that they contain about 90$\%$ of the counts in the protected regime.

The results are shown in Fig.\ref{fig:Figure3}. Starting from zero atoms, $\Fout$ increases with $N$, reaches a maximum for a measured coupling $\Omega/(2\pi) \simeq 300$ MHz corresponding to $N\simeq 25$ atoms, and then decreases to an asymptotic value. As expected, we find that the shape of $\Fout$ data is robust against the choice $\Delta f$ (see Fig. \ref{fig:comparing_exclusion_window}) and closely traces the calculated $S_{PW}$ (see Methods). This allows us to interpret the common behavior of $\Fout$ and $S_{PW}$ in terms of two competing effects. As $N$ increases, the dimension ($N+1$) of the Hilbert space of the system rises, and so does the number of states available to carry part of the photonic excitation outside $\Delta f$. The collective coupling $\Omega$ also increases (scaling as $\sqrt{N}$), and spreads the $PW$s to eigenstates close in frequency of $\Omega$ as long as the system stays in the unprotected regime $\Omega \ll \Delta \omega$. Then, above a certain value of the collective coupling $\Omega_{th}$, the cavity protection effect starts to reduce $S_{PW}$ and $\Fout$. In the large collective coupling limit, $S_{PW}$  and $\Fout$ are low, because the photonic weight - and thus the photon counts - concentrate in the two polaritonic states. To underscore the role of cavity protection contribution, we simulate $S_{PW}$ for a Lorentzian distribution with the same average frequency and width as the experimental one. In this case (see in Fig. \ref{fig:Figure3}), after the initial increase, $S_{PW}$  stays on a plateau as no cavity protection occurs. For our system, we note that the protection effect starts to play a role for a mesoscopic number of a few tens of atoms, which is several orders of magnitude lower than the typical number of emitters involved in solid-state based systems \cite{Putz2014a, Zhong2017b}. This allows us to explore this effect for a finite-sized system, and to directly observe the $PW$ concentration in the two polaritons.

\section{Spectral Engineering of Polaritons}

The coherence of the polaritons being preserved by the cavity protection, we can then harness the large sensitivity of the light-shifted atomic frequency to the trapping power to efficiently modulate the polaritonic frequencies. The intracavity dipole power is modulated by coupling two different frequencies of the dipole light into the cavity. The resulting beating  at a frequency $\omega_m/2 \pi$ leads to a modulation of the dipole potential and thus of the  average light-shifted frequency of the atoms $\overline{\omega_A}(t)=\overline{\omega_A^0}+ \beta_o \omega_m \cos(\omega_m t)$ where $\beta_o$ is the modulation index and $\overline{\omega_A^0}$ the average atomic frequency without modulation. The modulation frequency $\omega_m/2 \pi \sim 125\,$MHz is set to be larger than the width of the polaritons but smaller than the collective coupling $\Omega$. We record the transmission spectrum following the same procedure as before and average about 200 spectra. The results are presented in Fig. \ref{fig:Figure4}. Instead of the usual polariton doublet, the spectrum features two combs each consisting of several peaks. The frequency splitting between the centers of the two combs is given by the collective coupling $\Omega/2 \pi$ whereas the comb teeth are separated by the modulation frequency $\omega_m/2 \pi$. In contrast to the experiment of Ref. \cite{Clark2019}  where $\omega_m \gg \Omega$, the transmission spectrum does not result from the coupling between the cavity field and a single-frequency  atomic excitation  given by one sideband of the modulated atomic transition. In our case, the photonic excitation couples to a multi-frequency atomic excitation, yielding polaritons featuring multiple frequencies in their spectrum.

For $\omega_m \ll \Omega$, the theoretical cavity spectrum is well approximated by  (see Supplementary Information) 
\begin{eqnarray}
S(\omega)  \propto   \sum_n   
 \frac{ J_n^2   \left(\beta_o/2 \right)}{(\omega-\omega_0-n\omega_m+ \Omega )^2 + \gamma_s^2} \\ \nonumber
+ \frac{ J_n^2   \left(\beta_o/2 \right)}{(\omega-\omega_0-n\omega_m - \Omega)^2+ \gamma_s^2} 
\end{eqnarray}\label{cavity_spectrum_exper_inter}
where $\omega_0 = \omega_c = \overline{\omega_A^0}$ and $\gamma_s=\frac{\kappa+\gamma}{2}$. Based on this expression, we define a modulation index for the polaritons given by $\beta_p=\beta_o/2$. We have measured $\beta_p$ for different values of $\beta_o$ (see inset of Fig. \ref{fig:Figure4}), and find very good agreement with the expected slope. The spectrum of the polaritons can thus be directly controlled by tuning the modulation index and/or the modulation frequency. This is possible thanks to the large sensitivity of the excited state to the trapping power. An important remark is that the width of each peak in the multi-frequency polaritons is similar to the ones obtained in the non-modulated case and is much narrower than the atomic frequency distribution as we operate in the cavity protected regime.
A slight asymmetry between the right and left part of each comb is clearly visible on  Fig. \ref{fig:Figure4}. This asymmetry, stemming from the coupling between the two polaritons induced by the modulation, is well reproduced in the simulated spectrum obtained by numerically integrating the master equation (see Methods and Supplementary Information).

\bigskip

Polaritons with multi-frequency components and yet small linewidth created using this new technique can lead to applications in quantum network for spectral shaping of single-photon emission  \cite{Lukin2020a, Craiciu2021, Shlesinger2021}. More generally, the highly controlled experimental platform described here is ideally suited to study the role of inhomogeneities and disorder in strong light-matter interfaces. It has already allowed us to observe and quantitatively describe the transition between the protected and the unprotected regimes, yielding a better understanding of the role of dark states. It can be extended to generate a new kind of entangled dark states through collective dissipation useful for quantum sensing and quantum simulation, as recently proposed \cite{PineiroOrioli2022}, and to investigate the regime of higher excitation which has been little explored so far \cite{Lei2022}.
Moreover, when combined with addressing capabilities \cite{Barredo2016, Endres2016, Periwal2021a, Deist2022}, controllable  disorder as demonstrated here will allow the study of transport in disordered ensembles, with the perspective to quantum simulate the properties of engineered polaritonic materials \cite{Chavez2021, Dubail2022}.

\emph{Note:} While preparing this manuscript, we became aware of related results in the group of J.-P.~Brantut (EPFL Lausanne) \cite{Sauerwein2022}, and both groups decided to coordinate their submissions.
\vspace{0.1cm}

\begin{acknowledgments}
\textbf{Acknowledgments}
We thank G. Pupillo and J. Schachenmayer for discussion within the ANR project CLIMAQS. This project has received funding from:
Agence Nationale de la Recherche (ANR) (SAROCEMA project, ANR-14-CE32-0002);
European Research Council (ERC) under the European Union's Horizon 2020 research and innovation programme Grant agreement No 671133 (EQUEMI project). It has been supported by Region Ile-de-France in the framework of DIM SIRTEQ. S. Schwartz acknowledges funding from the European Union under the Marie Sklodowska Curie Individual Fellowship Programme H2020-MSCA-IF-2014 (project No. 658253)

\textbf{Authors' contributions}
M.B., F.F. and R.L. built the experimental setup, M.B., P.-A.B., S.S. and R.L. performed the measurements and analyzed the data. P.-A.B., S.S., F.F., J.R. and R.L. interpreted the results and wrote the manuscript with input from all authors. 

\textbf{Competing interests}
The authors declare no competing interests.

\textbf{Data and materials availability}
The data that support the findings of this study are available from the
corresponding authors upon reasonable request.
\end{acknowledgments}

\bibliography{Cavity_Protection}% common bib file

\begin{thebibliography}{58}%
\makeatletter
\providecommand \@ifxundefined [1]{%
 \@ifx{#1\undefined}
}%
\providecommand \@ifnum [1]{%
 \ifnum #1\expandafter \@firstoftwo
 \else \expandafter \@secondoftwo
 \fi
}%
\providecommand \@ifx [1]{%
 \ifx #1\expandafter \@firstoftwo
 \else \expandafter \@secondoftwo
 \fi
}%
\providecommand \natexlab [1]{#1}%
\providecommand \enquote  [1]{``#1''}%
\providecommand \bibnamefont  [1]{#1}%
\providecommand \bibfnamefont [1]{#1}%
\providecommand \citenamefont [1]{#1}%
\providecommand \href@noop [0]{\@secondoftwo}%
\providecommand \href [0]{\begingroup \@sanitize@url \@href}%
\providecommand \@href[1]{\@@startlink{#1}\@@href}%
\providecommand \@@href[1]{\endgroup#1\@@endlink}%
\providecommand \@sanitize@url [0]{\catcode `\\12\catcode `\$12\catcode
  `\&12\catcode `\#12\catcode `\^12\catcode `\_12\catcode `\%12\relax}%
\providecommand \@@startlink[1]{}%
\providecommand \@@endlink[0]{}%
\providecommand \url  [0]{\begingroup\@sanitize@url \@url }%
\providecommand \@url [1]{\endgroup\@href {#1}{\urlprefix }}%
\providecommand \urlprefix  [0]{URL }%
\providecommand \Eprint [0]{\href }%
\providecommand \doibase [0]{https://doi.org/}%
\providecommand \selectlanguage [0]{\@gobble}%
\providecommand \bibinfo  [0]{\@secondoftwo}%
\providecommand \bibfield  [0]{\@secondoftwo}%
\providecommand \translation [1]{[#1]}%
\providecommand \BibitemOpen [0]{}%
\providecommand \bibitemStop [0]{}%
\providecommand \bibitemNoStop [0]{.\EOS\space}%
\providecommand \EOS [0]{\spacefactor3000\relax}%
\providecommand \BibitemShut  [1]{\csname bibitem#1\endcsname}%
\let\auto@bib@innerbib\@empty
%</preamble>
\bibitem [{\citenamefont {Kimble}(1998)}]{Kimble1998}%
  \BibitemOpen
  \bibfield  {author} {\bibinfo {author} {\bibfnamefont {H.~J.}\ \bibnamefont
  {Kimble}},\ }\bibfield  {title} {\bibinfo {title} {Strong interactions of
  single atoms and photons in cavity {{QED}}},\ }\href
  {https://doi.org/10.1238/Physica.Topical.076a00127} {\bibfield  {journal}
  {\bibinfo  {journal} {Phys. Scr.}\ }\textbf {\bibinfo {volume} {1998}},\
  \bibinfo {pages} {127} (\bibinfo {year} {1998})}\BibitemShut {NoStop}%
\bibitem [{\citenamefont {Haroche}\ and\ \citenamefont
  {Raimond}(2006)}]{Haroche2006}%
  \BibitemOpen
  \bibfield  {author} {\bibinfo {author} {\bibfnamefont {S.}~\bibnamefont
  {Haroche}}\ and\ \bibinfo {author} {\bibfnamefont {J.-M.}\ \bibnamefont
  {Raimond}},\ }\href@noop {} {\emph {\bibinfo {title} {Exploring the
  {{Quantum}}: {{Atoms}}, {{Cavities}}, and {{Photons}}}}}\ (\bibinfo
  {publisher} {{Oxford University Press}},\ \bibinfo {year} {2006})\BibitemShut
  {NoStop}%
\bibitem [{\citenamefont {Solnyshkov}\ \emph {et~al.}(2021)\citenamefont
  {Solnyshkov}, \citenamefont {Malpuech}, \citenamefont {{St-Jean}},
  \citenamefont {Ravets}, \citenamefont {Bloch},\ and\ \citenamefont
  {Amo}}]{Solnyshkov2021}%
  \BibitemOpen
  \bibfield  {author} {\bibinfo {author} {\bibfnamefont {D.~D.}\ \bibnamefont
  {Solnyshkov}}, \bibinfo {author} {\bibfnamefont {G.}~\bibnamefont
  {Malpuech}}, \bibinfo {author} {\bibfnamefont {P.}~\bibnamefont {{St-Jean}}},
  \bibinfo {author} {\bibfnamefont {S.}~\bibnamefont {Ravets}}, \bibinfo
  {author} {\bibfnamefont {J.}~\bibnamefont {Bloch}},\ and\ \bibinfo {author}
  {\bibfnamefont {A.}~\bibnamefont {Amo}},\ }\bibfield  {title} {\bibinfo
  {title} {Microcavity polaritons for topological photonics [{{Invited}}]},\
  }\href {https://doi.org/10.1364/OME.414890} {\bibfield  {journal} {\bibinfo
  {journal} {Opt. Mater. Express}\ }\textbf {\bibinfo {volume} {11}},\ \bibinfo
  {pages} {1119} (\bibinfo {year} {2021})}\BibitemShut {NoStop}%
\bibitem [{\citenamefont {Blais}\ \emph {et~al.}(2020)\citenamefont {Blais},
  \citenamefont {Girvin},\ and\ \citenamefont {Oliver}}]{Blais2020b}%
  \BibitemOpen
  \bibfield  {author} {\bibinfo {author} {\bibfnamefont {A.}~\bibnamefont
  {Blais}}, \bibinfo {author} {\bibfnamefont {S.~M.}\ \bibnamefont {Girvin}},\
  and\ \bibinfo {author} {\bibfnamefont {W.~D.}\ \bibnamefont {Oliver}},\
  }\bibfield  {title} {\bibinfo {title} {Quantum information processing and
  quantum optics with circuit quantum electrodynamics},\ }\href
  {https://doi.org/10.1038/s41567-020-0806-z} {\bibfield  {journal} {\bibinfo
  {journal} {Nat. Phys.}\ }\textbf {\bibinfo {volume} {16}},\ \bibinfo {pages}
  {247} (\bibinfo {year} {2020})}\BibitemShut {NoStop}%
\bibitem [{\citenamefont {Chang}\ \emph {et~al.}(2018)\citenamefont {Chang},
  \citenamefont {Douglas}, \citenamefont {{Gonz{\'a}lez-Tudela}}, \citenamefont
  {Hung},\ and\ \citenamefont {Kimble}}]{Chang2018a}%
  \BibitemOpen
  \bibfield  {author} {\bibinfo {author} {\bibfnamefont {D.~E.}\ \bibnamefont
  {Chang}}, \bibinfo {author} {\bibfnamefont {J.~S.}\ \bibnamefont {Douglas}},
  \bibinfo {author} {\bibfnamefont {A.}~\bibnamefont {{Gonz{\'a}lez-Tudela}}},
  \bibinfo {author} {\bibfnamefont {C.-L.}\ \bibnamefont {Hung}},\ and\
  \bibinfo {author} {\bibfnamefont {H.~J.}\ \bibnamefont {Kimble}},\ }\bibfield
   {title} {\bibinfo {title} {Colloquium: {{Quantum}} matter built from
  nanoscopic lattices of atoms and photons},\ }\href
  {https://doi.org/10.1103/RevModPhys.90.031002} {\bibfield  {journal}
  {\bibinfo  {journal} {Rev. Mod. Phys.}\ }\textbf {\bibinfo {volume} {90}},\
  \bibinfo {pages} {031002} (\bibinfo {year} {2018})}\BibitemShut {NoStop}%
\bibitem [{\citenamefont {Ebbesen}(2016)}]{Ebbesen2016}%
  \BibitemOpen
  \bibfield  {author} {\bibinfo {author} {\bibfnamefont {T.~W.}\ \bibnamefont
  {Ebbesen}},\ }\bibfield  {title} {\bibinfo {title} {Hybrid light-matter
  states in a molecular and material science perspective},\ }\href
  {https://doi.org/10.1021/acs.accounts.6b00295} {\bibfield  {journal}
  {\bibinfo  {journal} {ACCOUNTS OF CHEMICAL RESEARCH}\ }\textbf {\bibinfo
  {volume} {49}},\ \bibinfo {pages} {2403} (\bibinfo {year}
  {2016})}\BibitemShut {NoStop}%
\bibitem [{\citenamefont {Bussi{\`e}res}\ \emph {et~al.}(2013)\citenamefont
  {Bussi{\`e}res}, \citenamefont {Sangouard}, \citenamefont {Afzelius},
  \citenamefont {de~Riedmatten}, \citenamefont {Simon},\ and\ \citenamefont
  {Tittel}}]{Bussieres2013}%
  \BibitemOpen
  \bibfield  {author} {\bibinfo {author} {\bibfnamefont {F.}~\bibnamefont
  {Bussi{\`e}res}}, \bibinfo {author} {\bibfnamefont {N.}~\bibnamefont
  {Sangouard}}, \bibinfo {author} {\bibfnamefont {M.}~\bibnamefont {Afzelius}},
  \bibinfo {author} {\bibfnamefont {H.}~\bibnamefont {de~Riedmatten}}, \bibinfo
  {author} {\bibfnamefont {C.}~\bibnamefont {Simon}},\ and\ \bibinfo {author}
  {\bibfnamefont {W.}~\bibnamefont {Tittel}},\ }\bibfield  {title} {\bibinfo
  {title} {Prospective applications of optical quantum memories},\ }\href
  {https://doi.org/10.1080/09500340.2013.856482} {\bibfield  {journal}
  {\bibinfo  {journal} {Journal of Modern Optics}\ }\textbf {\bibinfo {volume}
  {60}},\ \bibinfo {pages} {1519} (\bibinfo {year} {2013})}\BibitemShut
  {NoStop}%
\bibitem [{\citenamefont {Heshami}\ \emph {et~al.}(2016)\citenamefont
  {Heshami}, \citenamefont {England}, \citenamefont {Humphreys}, \citenamefont
  {Bustard}, \citenamefont {Acosta}, \citenamefont {Nunn},\ and\ \citenamefont
  {Sussman}}]{Heshami2016a}%
  \BibitemOpen
  \bibfield  {author} {\bibinfo {author} {\bibfnamefont {K.}~\bibnamefont
  {Heshami}}, \bibinfo {author} {\bibfnamefont {D.~G.}\ \bibnamefont
  {England}}, \bibinfo {author} {\bibfnamefont {P.~C.}\ \bibnamefont
  {Humphreys}}, \bibinfo {author} {\bibfnamefont {P.~J.}\ \bibnamefont
  {Bustard}}, \bibinfo {author} {\bibfnamefont {V.~M.}\ \bibnamefont {Acosta}},
  \bibinfo {author} {\bibfnamefont {J.}~\bibnamefont {Nunn}},\ and\ \bibinfo
  {author} {\bibfnamefont {B.~J.}\ \bibnamefont {Sussman}},\ }\bibfield
  {title} {\bibinfo {title} {Quantum memories: Emerging applications and recent
  advances},\ }\href {https://doi.org/10.1080/09500340.2016.1148212} {\bibfield
   {journal} {\bibinfo  {journal} {J Mod Opt}\ }\textbf {\bibinfo {volume}
  {63}},\ \bibinfo {pages} {2005} (\bibinfo {year} {2016})}\BibitemShut
  {NoStop}%
\bibitem [{\citenamefont {Zhong}\ \emph {et~al.}(2017)\citenamefont {Zhong},
  \citenamefont {Kindem}, \citenamefont {Rochman},\ and\ \citenamefont
  {Faraon}}]{Zhong2017b}%
  \BibitemOpen
  \bibfield  {author} {\bibinfo {author} {\bibfnamefont {T.}~\bibnamefont
  {Zhong}}, \bibinfo {author} {\bibfnamefont {J.~M.}\ \bibnamefont {Kindem}},
  \bibinfo {author} {\bibfnamefont {J.}~\bibnamefont {Rochman}},\ and\ \bibinfo
  {author} {\bibfnamefont {A.}~\bibnamefont {Faraon}},\ }\bibfield  {title}
  {\bibinfo {title} {Interfacing broadband photonic qubits to on-chip
  cavity-protected rare-earth ensembles},\ }\href
  {https://doi.org/10.1038/ncomms14107} {\bibfield  {journal} {\bibinfo
  {journal} {Nat Commun}\ }\textbf {\bibinfo {volume} {8}},\ \bibinfo {pages}
  {14107} (\bibinfo {year} {2017})}\BibitemShut {NoStop}%
\bibitem [{\citenamefont {Haas}\ \emph {et~al.}(2014)\citenamefont {Haas},
  \citenamefont {Volz}, \citenamefont {Gehr}, \citenamefont {Reichel},\ and\
  \citenamefont {Esteve}}]{Haas2014}%
  \BibitemOpen
  \bibfield  {author} {\bibinfo {author} {\bibfnamefont {F.}~\bibnamefont
  {Haas}}, \bibinfo {author} {\bibfnamefont {J.}~\bibnamefont {Volz}}, \bibinfo
  {author} {\bibfnamefont {R.}~\bibnamefont {Gehr}}, \bibinfo {author}
  {\bibfnamefont {J.}~\bibnamefont {Reichel}},\ and\ \bibinfo {author}
  {\bibfnamefont {J.}~\bibnamefont {Esteve}},\ }\bibfield  {title} {\bibinfo
  {title} {Entangled {{States}} of {{More Than}} 40 {{Atoms}} in an {{Optical
  Fiber Cavity}}},\ }\href {https://doi.org/10.1126/science.1248905} {\bibfield
   {journal} {\bibinfo  {journal} {Science}\ }\textbf {\bibinfo {volume}
  {344}},\ \bibinfo {pages} {180} (\bibinfo {year} {2014})}\BibitemShut
  {NoStop}%
\bibitem [{\citenamefont {McConnell}\ \emph {et~al.}(2015)\citenamefont
  {McConnell}, \citenamefont {Zhang}, \citenamefont {Hu}, \citenamefont
  {{\'C}uk},\ and\ \citenamefont {Vuleti{\'c}}}]{Mcconnell2015}%
  \BibitemOpen
  \bibfield  {author} {\bibinfo {author} {\bibfnamefont {R.}~\bibnamefont
  {McConnell}}, \bibinfo {author} {\bibfnamefont {H.}~\bibnamefont {Zhang}},
  \bibinfo {author} {\bibfnamefont {J.}~\bibnamefont {Hu}}, \bibinfo {author}
  {\bibfnamefont {S.}~\bibnamefont {{\'C}uk}},\ and\ \bibinfo {author}
  {\bibfnamefont {V.}~\bibnamefont {Vuleti{\'c}}},\ }\bibfield  {title}
  {\bibinfo {title} {Entanglement with negative {{Wigner}} function of almost
  3,000 atoms heralded by one photon},\ }\href
  {https://doi.org/10.1038/nature14293} {\bibfield  {journal} {\bibinfo
  {journal} {Nature}\ }\textbf {\bibinfo {volume} {519}},\ \bibinfo {pages}
  {439} (\bibinfo {year} {2015})}\BibitemShut {NoStop}%
\bibitem [{\citenamefont {Welte}\ \emph {et~al.}(2017)\citenamefont {Welte},
  \citenamefont {Hacker}, \citenamefont {Daiss}, \citenamefont {Ritter},\ and\
  \citenamefont {Rempe}}]{Welte2017a}%
  \BibitemOpen
  \bibfield  {author} {\bibinfo {author} {\bibfnamefont {S.}~\bibnamefont
  {Welte}}, \bibinfo {author} {\bibfnamefont {B.}~\bibnamefont {Hacker}},
  \bibinfo {author} {\bibfnamefont {S.}~\bibnamefont {Daiss}}, \bibinfo
  {author} {\bibfnamefont {S.}~\bibnamefont {Ritter}},\ and\ \bibinfo {author}
  {\bibfnamefont {G.}~\bibnamefont {Rempe}},\ }\bibfield  {title} {\bibinfo
  {title} {Cavity {{Carving}} of {{Atomic Bell States}}},\ }\href
  {https://doi.org/10.1103/PhysRevLett.118.210503} {\bibfield  {journal}
  {\bibinfo  {journal} {Phys. Rev. Lett.}\ }\textbf {\bibinfo {volume} {118}},\
  \bibinfo {pages} {210503} (\bibinfo {year} {2017})}\BibitemShut {NoStop}%
\bibitem [{\citenamefont {Leroux}\ \emph {et~al.}(2010)\citenamefont {Leroux},
  \citenamefont {{Schleier-Smith}},\ and\ \citenamefont
  {Vuleti{\'c}}}]{Leroux2010}%
  \BibitemOpen
  \bibfield  {author} {\bibinfo {author} {\bibfnamefont {I.~D.}\ \bibnamefont
  {Leroux}}, \bibinfo {author} {\bibfnamefont {M.~H.}\ \bibnamefont
  {{Schleier-Smith}}},\ and\ \bibinfo {author} {\bibfnamefont {V.}~\bibnamefont
  {Vuleti{\'c}}},\ }\bibfield  {title} {\bibinfo {title} {Implementation of
  {{Cavity Squeezing}} of a {{Collective Atomic Spin}}},\ }\href
  {https://doi.org/10.1103/PhysRevLett.104.073602} {\bibfield  {journal}
  {\bibinfo  {journal} {Phys. Rev. Lett.}\ }\textbf {\bibinfo {volume} {104}},\
  \bibinfo {pages} {073602} (\bibinfo {year} {2010})}\BibitemShut {NoStop}%
\bibitem [{\citenamefont {{Schleier-Smith}}\ \emph {et~al.}(2010)\citenamefont
  {{Schleier-Smith}}, \citenamefont {Leroux},\ and\ \citenamefont
  {Vuleti{\'c}}}]{Schleier-smith2010}%
  \BibitemOpen
  \bibfield  {author} {\bibinfo {author} {\bibfnamefont {M.~H.}\ \bibnamefont
  {{Schleier-Smith}}}, \bibinfo {author} {\bibfnamefont {I.~D.}\ \bibnamefont
  {Leroux}},\ and\ \bibinfo {author} {\bibfnamefont {V.}~\bibnamefont
  {Vuleti{\'c}}},\ }\bibfield  {title} {\bibinfo {title} {States of an
  {{Ensemble}} of {{Two-Level Atoms}} with {{Reduced Quantum Uncertainty}}},\
  }\href {https://doi.org/10.1103/PhysRevLett.104.073604} {\bibfield  {journal}
  {\bibinfo  {journal} {Phys. Rev. Lett.}\ }\textbf {\bibinfo {volume} {104}},\
  \bibinfo {pages} {073604} (\bibinfo {year} {2010})}\BibitemShut {NoStop}%
\bibitem [{\citenamefont {Ma}\ \emph {et~al.}(2011)\citenamefont {Ma},
  \citenamefont {Wang}, \citenamefont {Sun},\ and\ \citenamefont
  {Nori}}]{Ma2011}%
  \BibitemOpen
  \bibfield  {author} {\bibinfo {author} {\bibfnamefont {J.}~\bibnamefont
  {Ma}}, \bibinfo {author} {\bibfnamefont {X.}~\bibnamefont {Wang}}, \bibinfo
  {author} {\bibfnamefont {C.~P.}\ \bibnamefont {Sun}},\ and\ \bibinfo {author}
  {\bibfnamefont {F.}~\bibnamefont {Nori}},\ }\bibfield  {title} {\bibinfo
  {title} {Quantum spin squeezing},\ }\href
  {https://doi.org/10.1016/j.physrep.2011.08.003} {\bibfield  {journal}
  {\bibinfo  {journal} {Physics Reports}\ }\textbf {\bibinfo {volume} {509}},\
  \bibinfo {pages} {89} (\bibinfo {year} {2011})}\BibitemShut {NoStop}%
\bibitem [{\citenamefont {Hosten}\ \emph {et~al.}(2016)\citenamefont {Hosten},
  \citenamefont {Engelsen}, \citenamefont {Krishnakumar},\ and\ \citenamefont
  {Kasevich}}]{Hosten2016}%
  \BibitemOpen
  \bibfield  {author} {\bibinfo {author} {\bibfnamefont {O.}~\bibnamefont
  {Hosten}}, \bibinfo {author} {\bibfnamefont {N.~J.}\ \bibnamefont
  {Engelsen}}, \bibinfo {author} {\bibfnamefont {R.}~\bibnamefont
  {Krishnakumar}},\ and\ \bibinfo {author} {\bibfnamefont {M.~A.}\ \bibnamefont
  {Kasevich}},\ }\bibfield  {title} {\bibinfo {title} {Measurement noise 100
  times lower than the quantum-projection limit using entangled atoms},\ }\href
  {https://doi.org/10.1038/nature16176} {\bibfield  {journal} {\bibinfo
  {journal} {Nature}\ }\textbf {\bibinfo {volume} {529}},\ \bibinfo {pages}
  {505} (\bibinfo {year} {2016})}\BibitemShut {NoStop}%
\bibitem [{\citenamefont {Pezz{\`e}}\ \emph {et~al.}(2018)\citenamefont
  {Pezz{\`e}}, \citenamefont {Smerzi}, \citenamefont {Oberthaler},
  \citenamefont {Schmied},\ and\ \citenamefont {Treutlein}}]{Pezze2018}%
  \BibitemOpen
  \bibfield  {author} {\bibinfo {author} {\bibfnamefont {L.}~\bibnamefont
  {Pezz{\`e}}}, \bibinfo {author} {\bibfnamefont {A.}~\bibnamefont {Smerzi}},
  \bibinfo {author} {\bibfnamefont {M.~K.}\ \bibnamefont {Oberthaler}},
  \bibinfo {author} {\bibfnamefont {R.}~\bibnamefont {Schmied}},\ and\ \bibinfo
  {author} {\bibfnamefont {P.}~\bibnamefont {Treutlein}},\ }\bibfield  {title}
  {\bibinfo {title} {Quantum metrology with nonclassical states of atomic
  ensembles},\ }\href {https://doi.org/10.1103/RevModPhys.90.035005} {\bibfield
   {journal} {\bibinfo  {journal} {Rev. Mod. Phys.}\ }\textbf {\bibinfo
  {volume} {90}},\ \bibinfo {pages} {035005} (\bibinfo {year}
  {2018})}\BibitemShut {NoStop}%
\bibitem [{\citenamefont {Baumann}\ \emph {et~al.}(2010)\citenamefont
  {Baumann}, \citenamefont {Guerlin}, \citenamefont {Brennecke},\ and\
  \citenamefont {Esslinger}}]{Baumann2010a}%
  \BibitemOpen
  \bibfield  {author} {\bibinfo {author} {\bibfnamefont {K.}~\bibnamefont
  {Baumann}}, \bibinfo {author} {\bibfnamefont {C.}~\bibnamefont {Guerlin}},
  \bibinfo {author} {\bibfnamefont {F.}~\bibnamefont {Brennecke}},\ and\
  \bibinfo {author} {\bibfnamefont {T.}~\bibnamefont {Esslinger}},\ }\bibfield
  {title} {\bibinfo {title} {Dicke quantum phase transition with a superfluid
  gas in an optical cavity},\ }\href {https://doi.org/10.1038/nature09009}
  {\bibfield  {journal} {\bibinfo  {journal} {Nature}\ }\textbf {\bibinfo
  {volume} {464}},\ \bibinfo {pages} {1301} (\bibinfo {year}
  {2010})}\BibitemShut {NoStop}%
\bibitem [{\citenamefont {Kroeze}\ \emph {et~al.}(2018)\citenamefont {Kroeze},
  \citenamefont {Guo}, \citenamefont {Vaidya}, \citenamefont {Keeling},\ and\
  \citenamefont {Lev}}]{Kroeze2018}%
  \BibitemOpen
  \bibfield  {author} {\bibinfo {author} {\bibfnamefont {R.~M.}\ \bibnamefont
  {Kroeze}}, \bibinfo {author} {\bibfnamefont {Y.}~\bibnamefont {Guo}},
  \bibinfo {author} {\bibfnamefont {V.~D.}\ \bibnamefont {Vaidya}}, \bibinfo
  {author} {\bibfnamefont {J.}~\bibnamefont {Keeling}},\ and\ \bibinfo {author}
  {\bibfnamefont {B.~L.}\ \bibnamefont {Lev}},\ }\bibfield  {title} {\bibinfo
  {title} {Spinor {{Self-Ordering}} of a {{Quantum Gas}} in a {{Cavity}}},\
  }\href {https://doi.org/10.1103/PhysRevLett.121.163601} {\bibfield  {journal}
  {\bibinfo  {journal} {Phys. Rev. Lett.}\ }\textbf {\bibinfo {volume} {121}},\
  \bibinfo {pages} {163601} (\bibinfo {year} {2018})}\BibitemShut {NoStop}%
\bibitem [{\citenamefont {Muniz}\ \emph {et~al.}(2020)\citenamefont {Muniz},
  \citenamefont {Barberena}, \citenamefont {{Lewis-Swan}}, \citenamefont
  {Young}, \citenamefont {Cline}, \citenamefont {Rey},\ and\ \citenamefont
  {Thompson}}]{Muniz2020}%
  \BibitemOpen
  \bibfield  {author} {\bibinfo {author} {\bibfnamefont {J.~A.}\ \bibnamefont
  {Muniz}}, \bibinfo {author} {\bibfnamefont {D.}~\bibnamefont {Barberena}},
  \bibinfo {author} {\bibfnamefont {R.~J.}\ \bibnamefont {{Lewis-Swan}}},
  \bibinfo {author} {\bibfnamefont {D.~J.}\ \bibnamefont {Young}}, \bibinfo
  {author} {\bibfnamefont {J.~R.~K.}\ \bibnamefont {Cline}}, \bibinfo {author}
  {\bibfnamefont {A.~M.}\ \bibnamefont {Rey}},\ and\ \bibinfo {author}
  {\bibfnamefont {J.~K.}\ \bibnamefont {Thompson}},\ }\bibfield  {title}
  {\bibinfo {title} {Exploring dynamical phase transitions with cold atoms in
  an optical cavity},\ }\href {https://doi.org/10.1038/s41586-020-2224-x}
  {\bibfield  {journal} {\bibinfo  {journal} {Nature}\ }\textbf {\bibinfo
  {volume} {580}},\ \bibinfo {pages} {602} (\bibinfo {year}
  {2020})}\BibitemShut {NoStop}%
\bibitem [{\citenamefont {Periwal}\ \emph {et~al.}(2021)\citenamefont
  {Periwal}, \citenamefont {Cooper}, \citenamefont {Kunkel}, \citenamefont
  {Wienand}, \citenamefont {Davis},\ and\ \citenamefont
  {{Schleier-Smith}}}]{Periwal2021a}%
  \BibitemOpen
  \bibfield  {author} {\bibinfo {author} {\bibfnamefont {A.}~\bibnamefont
  {Periwal}}, \bibinfo {author} {\bibfnamefont {E.~S.}\ \bibnamefont {Cooper}},
  \bibinfo {author} {\bibfnamefont {P.}~\bibnamefont {Kunkel}}, \bibinfo
  {author} {\bibfnamefont {J.~F.}\ \bibnamefont {Wienand}}, \bibinfo {author}
  {\bibfnamefont {E.~J.}\ \bibnamefont {Davis}},\ and\ \bibinfo {author}
  {\bibfnamefont {M.}~\bibnamefont {{Schleier-Smith}}},\ }\bibfield  {title}
  {\bibinfo {title} {Programmable interactions and emergent geometry in an
  array of atom clouds},\ }\href {https://doi.org/10.1038/s41586-021-04156-0}
  {\bibfield  {journal} {\bibinfo  {journal} {Nature}\ }\textbf {\bibinfo
  {volume} {600}},\ \bibinfo {pages} {630} (\bibinfo {year}
  {2021})}\BibitemShut {NoStop}%
\bibitem [{\citenamefont {Basov}\ \emph {et~al.}(2020)\citenamefont {Basov},
  \citenamefont {{Asenjo-Garcia}}, \citenamefont {Schuck}, \citenamefont
  {Zhu},\ and\ \citenamefont {Rubio}}]{Basov2020}%
  \BibitemOpen
  \bibfield  {author} {\bibinfo {author} {\bibfnamefont {D.~N.}\ \bibnamefont
  {Basov}}, \bibinfo {author} {\bibfnamefont {A.}~\bibnamefont
  {{Asenjo-Garcia}}}, \bibinfo {author} {\bibfnamefont {P.~J.}\ \bibnamefont
  {Schuck}}, \bibinfo {author} {\bibfnamefont {X.}~\bibnamefont {Zhu}},\ and\
  \bibinfo {author} {\bibfnamefont {A.}~\bibnamefont {Rubio}},\ }\bibfield
  {title} {\bibinfo {title} {Polariton panorama},\ }\href
  {https://doi.org/10.1515/nanoph-2020-0449} {\bibfield  {journal} {\bibinfo
  {journal} {Nanophotonics}\ }\textbf {\bibinfo {volume} {10}},\ \bibinfo
  {pages} {549} (\bibinfo {year} {2020})}\BibitemShut {NoStop}%
\bibitem [{\citenamefont {Tavis}\ and\ \citenamefont
  {Cummings}(1968)}]{Tavis1968}%
  \BibitemOpen
  \bibfield  {author} {\bibinfo {author} {\bibfnamefont {M.}~\bibnamefont
  {Tavis}}\ and\ \bibinfo {author} {\bibfnamefont {F.~W.}\ \bibnamefont
  {Cummings}},\ }\bibfield  {title} {\bibinfo {title} {Exact {{Solution}} for
  an \${{N}}\$-{{Molecule---Radiation-Field Hamiltonian}}},\ }\href
  {https://doi.org/10.1103/PhysRev.170.379} {\bibfield  {journal} {\bibinfo
  {journal} {Phys. Rev.}\ }\textbf {\bibinfo {volume} {170}},\ \bibinfo {pages}
  {379} (\bibinfo {year} {1968})}\BibitemShut {NoStop}%
\bibitem [{\citenamefont {Orgiu}\ \emph {et~al.}(2015)\citenamefont {Orgiu},
  \citenamefont {George}, \citenamefont {Hutchison}, \citenamefont {Devaux},
  \citenamefont {Dayen}, \citenamefont {Doudin}, \citenamefont {Stellacci},
  \citenamefont {Genet}, \citenamefont {Schachenmayer}, \citenamefont {Genes},
  \citenamefont {Pupillo}, \citenamefont {Samor{\`i}},\ and\ \citenamefont
  {Ebbesen}}]{Orgiu2015}%
  \BibitemOpen
  \bibfield  {author} {\bibinfo {author} {\bibfnamefont {E.}~\bibnamefont
  {Orgiu}}, \bibinfo {author} {\bibfnamefont {J.}~\bibnamefont {George}},
  \bibinfo {author} {\bibfnamefont {J.~A.}\ \bibnamefont {Hutchison}}, \bibinfo
  {author} {\bibfnamefont {E.}~\bibnamefont {Devaux}}, \bibinfo {author}
  {\bibfnamefont {J.~F.}\ \bibnamefont {Dayen}}, \bibinfo {author}
  {\bibfnamefont {B.}~\bibnamefont {Doudin}}, \bibinfo {author} {\bibfnamefont
  {F.}~\bibnamefont {Stellacci}}, \bibinfo {author} {\bibfnamefont
  {C.}~\bibnamefont {Genet}}, \bibinfo {author} {\bibfnamefont
  {J.}~\bibnamefont {Schachenmayer}}, \bibinfo {author} {\bibfnamefont
  {C.}~\bibnamefont {Genes}}, \bibinfo {author} {\bibfnamefont
  {G.}~\bibnamefont {Pupillo}}, \bibinfo {author} {\bibfnamefont
  {P.}~\bibnamefont {Samor{\`i}}},\ and\ \bibinfo {author} {\bibfnamefont
  {T.~W.}\ \bibnamefont {Ebbesen}},\ }\bibfield  {title} {\bibinfo {title}
  {Conductivity in organic semiconductors hybridized with the vacuum field},\
  }\href {https://doi.org/10.1038/nmat4392} {\bibfield  {journal} {\bibinfo
  {journal} {Nature Mater}\ }\textbf {\bibinfo {volume} {14}},\ \bibinfo
  {pages} {1123} (\bibinfo {year} {2015})}\BibitemShut {NoStop}%
\bibitem [{\citenamefont {Schachenmayer}\ \emph {et~al.}(2015)\citenamefont
  {Schachenmayer}, \citenamefont {Genes}, \citenamefont {Tignone},\ and\
  \citenamefont {Pupillo}}]{schachenmayer2015}%
  \BibitemOpen
  \bibfield  {author} {\bibinfo {author} {\bibfnamefont {J.}~\bibnamefont
  {Schachenmayer}}, \bibinfo {author} {\bibfnamefont {C.}~\bibnamefont
  {Genes}}, \bibinfo {author} {\bibfnamefont {E.}~\bibnamefont {Tignone}},\
  and\ \bibinfo {author} {\bibfnamefont {G.}~\bibnamefont {Pupillo}},\
  }\bibfield  {title} {\bibinfo {title} {Cavity-{{Enhanced Transport}} of
  {{Excitons}}},\ }\href {https://doi.org/10.1103/PhysRevLett.114.196403}
  {\bibfield  {journal} {\bibinfo  {journal} {Phys. Rev. Lett.}\ }\textbf
  {\bibinfo {volume} {114}},\ \bibinfo {pages} {196403} (\bibinfo {year}
  {2015})}\BibitemShut {NoStop}%
\bibitem [{\citenamefont {{Gonzalez-Ballestero}}\ \emph
  {et~al.}(2016)\citenamefont {{Gonzalez-Ballestero}}, \citenamefont {Feist},
  \citenamefont {Gonzalo~Bad{\'i}a}, \citenamefont {Moreno},\ and\
  \citenamefont {{Garcia-Vidal}}}]{Gonzalez-ballestero2016}%
  \BibitemOpen
  \bibfield  {author} {\bibinfo {author} {\bibfnamefont {C.}~\bibnamefont
  {{Gonzalez-Ballestero}}}, \bibinfo {author} {\bibfnamefont {J.}~\bibnamefont
  {Feist}}, \bibinfo {author} {\bibfnamefont {E.}~\bibnamefont
  {Gonzalo~Bad{\'i}a}}, \bibinfo {author} {\bibfnamefont {E.}~\bibnamefont
  {Moreno}},\ and\ \bibinfo {author} {\bibfnamefont {F.~J.}\ \bibnamefont
  {{Garcia-Vidal}}},\ }\bibfield  {title} {\bibinfo {title} {Uncoupled {{Dark
  States Can Inherit Polaritonic Properties}}},\ }\href
  {https://doi.org/10.1103/PhysRevLett.117.156402} {\bibfield  {journal}
  {\bibinfo  {journal} {Phys. Rev. Lett.}\ }\textbf {\bibinfo {volume} {117}},\
  \bibinfo {pages} {156402} (\bibinfo {year} {2016})}\BibitemShut {NoStop}%
\bibitem [{\citenamefont {Botzung}\ \emph {et~al.}(2020)\citenamefont
  {Botzung}, \citenamefont {Hagenm{\"u}ller}, \citenamefont {Sch{\"u}tz},
  \citenamefont {Dubail}, \citenamefont {Pupillo},\ and\ \citenamefont
  {Schachenmayer}}]{Botzung2020a}%
  \BibitemOpen
  \bibfield  {author} {\bibinfo {author} {\bibfnamefont {T.}~\bibnamefont
  {Botzung}}, \bibinfo {author} {\bibfnamefont {D.}~\bibnamefont
  {Hagenm{\"u}ller}}, \bibinfo {author} {\bibfnamefont {S.}~\bibnamefont
  {Sch{\"u}tz}}, \bibinfo {author} {\bibfnamefont {J.}~\bibnamefont {Dubail}},
  \bibinfo {author} {\bibfnamefont {G.}~\bibnamefont {Pupillo}},\ and\ \bibinfo
  {author} {\bibfnamefont {J.}~\bibnamefont {Schachenmayer}},\ }\bibfield
  {title} {\bibinfo {title} {Dark state semilocalization of quantum emitters in
  a cavity},\ }\href {https://doi.org/10.1103/PhysRevB.102.144202} {\bibfield
  {journal} {\bibinfo  {journal} {Phys. Rev. B}\ }\textbf {\bibinfo {volume}
  {102}},\ \bibinfo {pages} {144202} (\bibinfo {year} {2020})}\BibitemShut
  {NoStop}%
\bibitem [{\citenamefont {Ch{\'a}vez}\ \emph {et~al.}(2021)\citenamefont
  {Ch{\'a}vez}, \citenamefont {Mattiotti}, \citenamefont
  {{M{\'e}ndez-Berm{\'u}dez}}, \citenamefont {Borgonovi},\ and\ \citenamefont
  {Celardo}}]{Chavez2021}%
  \BibitemOpen
  \bibfield  {author} {\bibinfo {author} {\bibfnamefont {N.~C.}\ \bibnamefont
  {Ch{\'a}vez}}, \bibinfo {author} {\bibfnamefont {F.}~\bibnamefont
  {Mattiotti}}, \bibinfo {author} {\bibfnamefont {J.~A.}\ \bibnamefont
  {{M{\'e}ndez-Berm{\'u}dez}}}, \bibinfo {author} {\bibfnamefont
  {F.}~\bibnamefont {Borgonovi}},\ and\ \bibinfo {author} {\bibfnamefont
  {G.~L.}\ \bibnamefont {Celardo}},\ }\bibfield  {title} {\bibinfo {title}
  {Disorder-{{Enhanced}} and {{Disorder-Independent Transport}} with
  {{Long-Range Hopping}}: {{Application}} to {{Molecular Chains}} in {{Optical
  Cavities}}},\ }\href {https://doi.org/10.1103/PhysRevLett.126.153201}
  {\bibfield  {journal} {\bibinfo  {journal} {Phys. Rev. Lett.}\ }\textbf
  {\bibinfo {volume} {126}},\ \bibinfo {pages} {153201} (\bibinfo {year}
  {2021})}\BibitemShut {NoStop}%
\bibitem [{\citenamefont {Mivehvar}\ \emph {et~al.}(2021)\citenamefont
  {Mivehvar}, \citenamefont {Piazza}, \citenamefont {Donner},\ and\
  \citenamefont {Ritsch}}]{Mivehvar2021}%
  \BibitemOpen
  \bibfield  {author} {\bibinfo {author} {\bibfnamefont {F.}~\bibnamefont
  {Mivehvar}}, \bibinfo {author} {\bibfnamefont {F.}~\bibnamefont {Piazza}},
  \bibinfo {author} {\bibfnamefont {T.}~\bibnamefont {Donner}},\ and\ \bibinfo
  {author} {\bibfnamefont {H.}~\bibnamefont {Ritsch}},\ }\bibfield  {title}
  {\bibinfo {title} {Cavity {{QED}} with quantum gases: New paradigms in
  many-body physics},\ }\href {https://doi.org/10.1080/00018732.2021.1969727}
  {\bibfield  {journal} {\bibinfo  {journal} {Advances in Physics}\ }\textbf
  {\bibinfo {volume} {70}},\ \bibinfo {pages} {1} (\bibinfo {year}
  {2021})}\BibitemShut {NoStop}%
\bibitem [{\citenamefont {Houdr{\'e}}\ \emph {et~al.}(1996)\citenamefont
  {Houdr{\'e}}, \citenamefont {Stanley},\ and\ \citenamefont
  {Ilegems}}]{Houdre1996a}%
  \BibitemOpen
  \bibfield  {author} {\bibinfo {author} {\bibfnamefont {R.}~\bibnamefont
  {Houdr{\'e}}}, \bibinfo {author} {\bibfnamefont {R.~P.}\ \bibnamefont
  {Stanley}},\ and\ \bibinfo {author} {\bibfnamefont {M.}~\bibnamefont
  {Ilegems}},\ }\bibfield  {title} {\bibinfo {title} {Vacuum-field {{Rabi}}
  splitting in the presence of inhomogeneous broadening: {{Resolution}} of a
  homogeneous linewidth in an inhomogeneously broadened system},\ }\href
  {https://doi.org/10.1103/PhysRevA.53.2711} {\bibfield  {journal} {\bibinfo
  {journal} {Phys. Rev. A}\ }\textbf {\bibinfo {volume} {53}},\ \bibinfo
  {pages} {2711} (\bibinfo {year} {1996})}\BibitemShut {NoStop}%
\bibitem [{\citenamefont {Kurucz}\ \emph {et~al.}(2011)\citenamefont {Kurucz},
  \citenamefont {Wesenberg},\ and\ \citenamefont {M{\o}lmer}}]{Kurucz2011a}%
  \BibitemOpen
  \bibfield  {author} {\bibinfo {author} {\bibfnamefont {Z.}~\bibnamefont
  {Kurucz}}, \bibinfo {author} {\bibfnamefont {J.~H.}\ \bibnamefont
  {Wesenberg}},\ and\ \bibinfo {author} {\bibfnamefont {K.}~\bibnamefont
  {M{\o}lmer}},\ }\bibfield  {title} {\bibinfo {title} {Spectroscopic
  properties of inhomogeneously broadened spin ensembles in a cavity},\ }\href
  {https://doi.org/10.1103/PhysRevA.83.053852} {\bibfield  {journal} {\bibinfo
  {journal} {Phys. Rev. A}\ }\textbf {\bibinfo {volume} {83}},\ \bibinfo
  {pages} {053852} (\bibinfo {year} {2011})}\BibitemShut {NoStop}%
\bibitem [{\citenamefont {Diniz}\ \emph {et~al.}(2011)\citenamefont {Diniz},
  \citenamefont {Portolan}, \citenamefont {Ferreira}, \citenamefont
  {G{\'e}rard}, \citenamefont {Bertet},\ and\ \citenamefont
  {Auff{\`e}ves}}]{Diniz2011b}%
  \BibitemOpen
  \bibfield  {author} {\bibinfo {author} {\bibfnamefont {I.}~\bibnamefont
  {Diniz}}, \bibinfo {author} {\bibfnamefont {S.}~\bibnamefont {Portolan}},
  \bibinfo {author} {\bibfnamefont {R.}~\bibnamefont {Ferreira}}, \bibinfo
  {author} {\bibfnamefont {J.~M.}\ \bibnamefont {G{\'e}rard}}, \bibinfo
  {author} {\bibfnamefont {P.}~\bibnamefont {Bertet}},\ and\ \bibinfo {author}
  {\bibfnamefont {A.}~\bibnamefont {Auff{\`e}ves}},\ }\bibfield  {title}
  {\bibinfo {title} {Strongly coupling a cavity to inhomogeneous ensembles of
  emitters: {{Potential}} for long-lived solid-state quantum memories},\ }\href
  {https://doi.org/10.1103/PhysRevA.84.063810} {\bibfield  {journal} {\bibinfo
  {journal} {Phys. Rev. A}\ }\textbf {\bibinfo {volume} {84}},\ \bibinfo
  {pages} {063810} (\bibinfo {year} {2011})}\BibitemShut {NoStop}%
\bibitem [{\citenamefont {Putz}\ \emph {et~al.}(2014)\citenamefont {Putz},
  \citenamefont {Krimer}, \citenamefont {Ams{\"u}ss}, \citenamefont
  {Valookaran}, \citenamefont {N{\"o}bauer}, \citenamefont {Schmiedmayer},
  \citenamefont {Rotter},\ and\ \citenamefont {Majer}}]{Putz2014a}%
  \BibitemOpen
  \bibfield  {author} {\bibinfo {author} {\bibfnamefont {S.}~\bibnamefont
  {Putz}}, \bibinfo {author} {\bibfnamefont {D.~O.}\ \bibnamefont {Krimer}},
  \bibinfo {author} {\bibfnamefont {R.}~\bibnamefont {Ams{\"u}ss}}, \bibinfo
  {author} {\bibfnamefont {A.}~\bibnamefont {Valookaran}}, \bibinfo {author}
  {\bibfnamefont {T.}~\bibnamefont {N{\"o}bauer}}, \bibinfo {author}
  {\bibfnamefont {J.}~\bibnamefont {Schmiedmayer}}, \bibinfo {author}
  {\bibfnamefont {S.}~\bibnamefont {Rotter}},\ and\ \bibinfo {author}
  {\bibfnamefont {J.}~\bibnamefont {Majer}},\ }\bibfield  {title} {\bibinfo
  {title} {Protecting a spin ensemble against decoherence in the
  strong-coupling regime of cavity {{QED}}},\ }\href
  {https://doi.org/10.1038/nphys3050} {\bibfield  {journal} {\bibinfo
  {journal} {Nature Phys}\ }\textbf {\bibinfo {volume} {10}},\ \bibinfo {pages}
  {720} (\bibinfo {year} {2014})}\BibitemShut {NoStop}%
\bibitem [{\citenamefont {Breeze}\ \emph {et~al.}(2017)\citenamefont {Breeze},
  \citenamefont {Salvadori}, \citenamefont {Sathian}, \citenamefont {Alford},\
  and\ \citenamefont {Kay}}]{Breeze2017}%
  \BibitemOpen
  \bibfield  {author} {\bibinfo {author} {\bibfnamefont {J.~D.}\ \bibnamefont
  {Breeze}}, \bibinfo {author} {\bibfnamefont {E.}~\bibnamefont {Salvadori}},
  \bibinfo {author} {\bibfnamefont {J.}~\bibnamefont {Sathian}}, \bibinfo
  {author} {\bibfnamefont {N.~M.}\ \bibnamefont {Alford}},\ and\ \bibinfo
  {author} {\bibfnamefont {C.~W.~M.}\ \bibnamefont {Kay}},\ }\bibfield  {title}
  {\bibinfo {title} {Room-temperature cavity quantum electrodynamics with
  strongly coupled {{Dicke}} states},\ }\href
  {https://doi.org/10.1038/s41534-017-0041-3} {\bibfield  {journal} {\bibinfo
  {journal} {npj Quantum Inf}\ }\textbf {\bibinfo {volume} {3}},\ \bibinfo
  {pages} {40} (\bibinfo {year} {2017})}\BibitemShut {NoStop}%
\bibitem [{\citenamefont {Lukin}\ \emph {et~al.}(2020)\citenamefont {Lukin},
  \citenamefont {White}, \citenamefont {Trivedi}, \citenamefont {Guidry},
  \citenamefont {Morioka}, \citenamefont {Babin}, \citenamefont {Soykal},
  \citenamefont {{Ul-Hassan}}, \citenamefont {Son}, \citenamefont {Ohshima},
  \citenamefont {Vasireddy}, \citenamefont {Nasr}, \citenamefont {Sun},
  \citenamefont {MacLean}, \citenamefont {Dory}, \citenamefont {Nanni},
  \citenamefont {Wrachtrup}, \citenamefont {Kaiser},\ and\ \citenamefont {Vu{\v
  c}kovi{\'c}}}]{Lukin2020a}%
  \BibitemOpen
  \bibfield  {author} {\bibinfo {author} {\bibfnamefont {D.~M.}\ \bibnamefont
  {Lukin}}, \bibinfo {author} {\bibfnamefont {A.~D.}\ \bibnamefont {White}},
  \bibinfo {author} {\bibfnamefont {R.}~\bibnamefont {Trivedi}}, \bibinfo
  {author} {\bibfnamefont {M.~A.}\ \bibnamefont {Guidry}}, \bibinfo {author}
  {\bibfnamefont {N.}~\bibnamefont {Morioka}}, \bibinfo {author} {\bibfnamefont
  {C.}~\bibnamefont {Babin}}, \bibinfo {author} {\bibfnamefont {{\"O}.~O.}\
  \bibnamefont {Soykal}}, \bibinfo {author} {\bibfnamefont {J.}~\bibnamefont
  {{Ul-Hassan}}}, \bibinfo {author} {\bibfnamefont {N.~T.}\ \bibnamefont
  {Son}}, \bibinfo {author} {\bibfnamefont {T.}~\bibnamefont {Ohshima}},
  \bibinfo {author} {\bibfnamefont {P.~K.}\ \bibnamefont {Vasireddy}}, \bibinfo
  {author} {\bibfnamefont {M.~H.}\ \bibnamefont {Nasr}}, \bibinfo {author}
  {\bibfnamefont {S.}~\bibnamefont {Sun}}, \bibinfo {author} {\bibfnamefont
  {J.-P.~W.}\ \bibnamefont {MacLean}}, \bibinfo {author} {\bibfnamefont
  {C.}~\bibnamefont {Dory}}, \bibinfo {author} {\bibfnamefont {E.~A.}\
  \bibnamefont {Nanni}}, \bibinfo {author} {\bibfnamefont {J.}~\bibnamefont
  {Wrachtrup}}, \bibinfo {author} {\bibfnamefont {F.}~\bibnamefont {Kaiser}},\
  and\ \bibinfo {author} {\bibfnamefont {J.}~\bibnamefont {Vu{\v
  c}kovi{\'c}}},\ }\bibfield  {title} {\bibinfo {title} {Spectrally
  reconfigurable quantum emitters enabled by optimized fast modulation},\
  }\href {https://doi.org/10.1038/s41534-020-00310-0} {\bibfield  {journal}
  {\bibinfo  {journal} {npj Quantum Information}\ }\textbf {\bibinfo {volume}
  {6}},\ \bibinfo {pages} {1} (\bibinfo {year} {2020})}\BibitemShut {NoStop}%
\bibitem [{\citenamefont {Craiciu}\ \emph {et~al.}(2021)\citenamefont
  {Craiciu}, \citenamefont {Craiciu}, \citenamefont {Craiciu}, \citenamefont
  {Lei}, \citenamefont {Lei}, \citenamefont {Rochman}, \citenamefont {Rochman},
  \citenamefont {Bartholomew}, \citenamefont {Bartholomew}, \citenamefont
  {Bartholomew}, \citenamefont {Bartholomew}, \citenamefont {Faraon},\ and\
  \citenamefont {Faraon}}]{Craiciu2021}%
  \BibitemOpen
  \bibfield  {author} {\bibinfo {author} {\bibfnamefont {I.}~\bibnamefont
  {Craiciu}}, \bibinfo {author} {\bibfnamefont {I.}~\bibnamefont {Craiciu}},
  \bibinfo {author} {\bibfnamefont {I.}~\bibnamefont {Craiciu}}, \bibinfo
  {author} {\bibfnamefont {M.}~\bibnamefont {Lei}}, \bibinfo {author}
  {\bibfnamefont {M.}~\bibnamefont {Lei}}, \bibinfo {author} {\bibfnamefont
  {J.}~\bibnamefont {Rochman}}, \bibinfo {author} {\bibfnamefont
  {J.}~\bibnamefont {Rochman}}, \bibinfo {author} {\bibfnamefont {J.~G.}\
  \bibnamefont {Bartholomew}}, \bibinfo {author} {\bibfnamefont {J.~G.}\
  \bibnamefont {Bartholomew}}, \bibinfo {author} {\bibfnamefont {J.~G.}\
  \bibnamefont {Bartholomew}}, \bibinfo {author} {\bibfnamefont {J.~G.}\
  \bibnamefont {Bartholomew}}, \bibinfo {author} {\bibfnamefont
  {A.}~\bibnamefont {Faraon}},\ and\ \bibinfo {author} {\bibfnamefont
  {A.}~\bibnamefont {Faraon}},\ }\bibfield  {title} {\bibinfo {title}
  {Multifunctional on-chip storage at telecommunication wavelength for quantum
  networks},\ }\href {https://doi.org/10.1364/OPTICA.412211} {\bibfield
  {journal} {\bibinfo  {journal} {Optica, OPTICA}\ }\textbf {\bibinfo {volume}
  {8}},\ \bibinfo {pages} {114} (\bibinfo {year} {2021})}\BibitemShut {NoStop}%
\bibitem [{\citenamefont {Mishra}\ \emph {et~al.}(2021)\citenamefont {Mishra},
  \citenamefont {Trivedi}, \citenamefont {{Safavi-Naeini}},\ and\ \citenamefont
  {Vu{\v c}kovi{\'c}}}]{Mishra2021a}%
  \BibitemOpen
  \bibfield  {author} {\bibinfo {author} {\bibfnamefont {S.~D.}\ \bibnamefont
  {Mishra}}, \bibinfo {author} {\bibfnamefont {R.}~\bibnamefont {Trivedi}},
  \bibinfo {author} {\bibfnamefont {A.~H.}\ \bibnamefont {{Safavi-Naeini}}},\
  and\ \bibinfo {author} {\bibfnamefont {J.}~\bibnamefont {Vu{\v
  c}kovi{\'c}}},\ }\bibfield  {title} {\bibinfo {title} {Control {{Design}} for
  {{Inhomogeneous-Broadening Compensation}} in {{Single-Photon Transducers}}},\
  }\href {https://doi.org/10.1103/PhysRevApplied.16.044025} {\bibfield
  {journal} {\bibinfo  {journal} {Phys. Rev. Applied}\ }\textbf {\bibinfo
  {volume} {16}},\ \bibinfo {pages} {044025} (\bibinfo {year}
  {2021})}\BibitemShut {NoStop}%
\bibitem [{\citenamefont {White}\ \emph {et~al.}(2022)\citenamefont {White},
  \citenamefont {Trivedi}, \citenamefont {Narayanan},\ and\ \citenamefont
  {Vu{\v c}kovi{\'c}}}]{White2022}%
  \BibitemOpen
  \bibfield  {author} {\bibinfo {author} {\bibfnamefont {A.~D.}\ \bibnamefont
  {White}}, \bibinfo {author} {\bibfnamefont {R.}~\bibnamefont {Trivedi}},
  \bibinfo {author} {\bibfnamefont {K.}~\bibnamefont {Narayanan}},\ and\
  \bibinfo {author} {\bibfnamefont {J.}~\bibnamefont {Vu{\v c}kovi{\'c}}},\
  }\bibfield  {title} {\bibinfo {title} {Enhancing {{Superradiance}} in
  {{Spectrally Inhomogeneous Cavity QED Systems}} with {{Dynamic
  Modulation}}},\ }\href {https://doi.org/10.1021/acsphotonics.2c00581}
  {\bibfield  {journal} {\bibinfo  {journal} {ACS Photonics}\ }\textbf
  {\bibinfo {volume} {9}},\ \bibinfo {pages} {2467} (\bibinfo {year}
  {2022})}\BibitemShut {NoStop}%
\bibitem [{\citenamefont {Hunger}\ \emph {et~al.}(2010)\citenamefont {Hunger},
  \citenamefont {Steinmetz}, \citenamefont {Colombe}, \citenamefont {Deutsch},
  \citenamefont {H{\"a}nsch},\ and\ \citenamefont {Reichel}}]{Hunger2010}%
  \BibitemOpen
  \bibfield  {author} {\bibinfo {author} {\bibfnamefont {D.}~\bibnamefont
  {Hunger}}, \bibinfo {author} {\bibfnamefont {T.}~\bibnamefont {Steinmetz}},
  \bibinfo {author} {\bibfnamefont {Y.}~\bibnamefont {Colombe}}, \bibinfo
  {author} {\bibfnamefont {C.}~\bibnamefont {Deutsch}}, \bibinfo {author}
  {\bibfnamefont {T.~W.}\ \bibnamefont {H{\"a}nsch}},\ and\ \bibinfo {author}
  {\bibfnamefont {J.}~\bibnamefont {Reichel}},\ }\bibfield  {title} {\bibinfo
  {title} {A fiber {{Fabry}}\textendash{{Perot}} cavity with high finesse},\
  }\href {https://doi.org/10.1088/1367-2630/12/6/065038} {\bibfield  {journal}
  {\bibinfo  {journal} {New J. Phys.}\ }\textbf {\bibinfo {volume} {12}},\
  \bibinfo {pages} {065038} (\bibinfo {year} {2010})}\BibitemShut {NoStop}%
\bibitem [{\citenamefont {Garcia}\ \emph {et~al.}(2018)\citenamefont {Garcia},
  \citenamefont {Ferri}, \citenamefont {Ott}, \citenamefont {Reichel},\ and\
  \citenamefont {Long}}]{Garcia2018}%
  \BibitemOpen
  \bibfield  {author} {\bibinfo {author} {\bibfnamefont {S.}~\bibnamefont
  {Garcia}}, \bibinfo {author} {\bibfnamefont {F.}~\bibnamefont {Ferri}},
  \bibinfo {author} {\bibfnamefont {K.}~\bibnamefont {Ott}}, \bibinfo {author}
  {\bibfnamefont {J.}~\bibnamefont {Reichel}},\ and\ \bibinfo {author}
  {\bibfnamefont {R.}~\bibnamefont {Long}},\ }\bibfield  {title} {\bibinfo
  {title} {Dual-wavelength fiber {{Fabry-Perot}} cavities with engineered
  birefringence},\ }\href {https://doi.org/10.1364/OE.26.022249} {\bibfield
  {journal} {\bibinfo  {journal} {Opt. Express}\ }\textbf {\bibinfo {volume}
  {26}},\ \bibinfo {pages} {22249} (\bibinfo {year} {2018})}\BibitemShut
  {NoStop}%
\bibitem [{\citenamefont {Garcia}\ \emph {et~al.}(2020)\citenamefont {Garcia},
  \citenamefont {Ferri}, \citenamefont {Reichel},\ and\ \citenamefont
  {Long}}]{Garcia2020a}%
  \BibitemOpen
  \bibfield  {author} {\bibinfo {author} {\bibfnamefont {S.}~\bibnamefont
  {Garcia}}, \bibinfo {author} {\bibfnamefont {F.}~\bibnamefont {Ferri}},
  \bibinfo {author} {\bibfnamefont {J.}~\bibnamefont {Reichel}},\ and\ \bibinfo
  {author} {\bibfnamefont {R.}~\bibnamefont {Long}},\ }\bibfield  {title}
  {\bibinfo {title} {Overlapping two standing waves in a microcavity for a
  multi-atom photon interface},\ }\href {https://doi.org/10.1364/OE.392207}
  {\bibfield  {journal} {\bibinfo  {journal} {Opt. Express}\ }\textbf {\bibinfo
  {volume} {28}},\ \bibinfo {pages} {15515} (\bibinfo {year}
  {2020})}\BibitemShut {NoStop}%
\bibitem [{\citenamefont {Ferri}\ \emph {et~al.}(2020)\citenamefont {Ferri},
  \citenamefont {Garcia}, \citenamefont {Baghdad}, \citenamefont {Reichel},\
  and\ \citenamefont {Long}}]{Ferri2020a}%
  \BibitemOpen
  \bibfield  {author} {\bibinfo {author} {\bibfnamefont {F.}~\bibnamefont
  {Ferri}}, \bibinfo {author} {\bibfnamefont {S.}~\bibnamefont {Garcia}},
  \bibinfo {author} {\bibfnamefont {M.}~\bibnamefont {Baghdad}}, \bibinfo
  {author} {\bibfnamefont {J.}~\bibnamefont {Reichel}},\ and\ \bibinfo {author}
  {\bibfnamefont {R.}~\bibnamefont {Long}},\ }\bibfield  {title} {\bibinfo
  {title} {Mapping optical standing-waves of an open-access
  {{Fabry}}\textendash{{Perot}} cavity with a tapered fiber},\ }\href
  {https://doi.org/10.1063/1.5142709} {\bibfield  {journal} {\bibinfo
  {journal} {Review of Scientific Instruments}\ }\textbf {\bibinfo {volume}
  {91}},\ \bibinfo {pages} {033104} (\bibinfo {year} {2020})}\BibitemShut
  {NoStop}%
\bibitem [{\citenamefont {Brantut}\ \emph {et~al.}(2008)\citenamefont
  {Brantut}, \citenamefont {Cl{\'e}ment}, \citenamefont {{de Saint Vincent}},
  \citenamefont {Varoquaux}, \citenamefont {Nyman}, \citenamefont {Aspect},
  \citenamefont {Bourdel},\ and\ \citenamefont {Bouyer}}]{Brantut2008}%
  \BibitemOpen
  \bibfield  {author} {\bibinfo {author} {\bibfnamefont {J.~P.}\ \bibnamefont
  {Brantut}}, \bibinfo {author} {\bibfnamefont {J.~F.}\ \bibnamefont
  {Cl{\'e}ment}}, \bibinfo {author} {\bibfnamefont {M.~R.}\ \bibnamefont {{de
  Saint Vincent}}}, \bibinfo {author} {\bibfnamefont {G.}~\bibnamefont
  {Varoquaux}}, \bibinfo {author} {\bibfnamefont {R.~A.}\ \bibnamefont
  {Nyman}}, \bibinfo {author} {\bibfnamefont {A.}~\bibnamefont {Aspect}},
  \bibinfo {author} {\bibfnamefont {T.}~\bibnamefont {Bourdel}},\ and\ \bibinfo
  {author} {\bibfnamefont {P.}~\bibnamefont {Bouyer}},\ }\bibfield  {title}
  {\bibinfo {title} {Light-shift tomography in an optical-dipole trap for
  neutral atoms},\ }\href {https://doi.org/10.1103/PhysRevA.78.031401}
  {\bibfield  {journal} {\bibinfo  {journal} {Phys. Rev. A}\ }\textbf {\bibinfo
  {volume} {78}},\ \bibinfo {pages} {031401} (\bibinfo {year}
  {2008})}\BibitemShut {NoStop}%
\bibitem [{\citenamefont {Dubail}\ \emph {et~al.}(2022)\citenamefont {Dubail},
  \citenamefont {Botzung}, \citenamefont {Schachenmayer}, \citenamefont
  {Pupillo},\ and\ \citenamefont {Hagenm{\"u}ller}}]{Dubail2022}%
  \BibitemOpen
  \bibfield  {author} {\bibinfo {author} {\bibfnamefont {J.}~\bibnamefont
  {Dubail}}, \bibinfo {author} {\bibfnamefont {T.}~\bibnamefont {Botzung}},
  \bibinfo {author} {\bibfnamefont {J.}~\bibnamefont {Schachenmayer}}, \bibinfo
  {author} {\bibfnamefont {G.}~\bibnamefont {Pupillo}},\ and\ \bibinfo {author}
  {\bibfnamefont {D.}~\bibnamefont {Hagenm{\"u}ller}},\ }\bibfield  {title}
  {\bibinfo {title} {Large random arrowhead matrices: {{Multifractality}},
  semilocalization, and protected transport in disordered quantum spins coupled
  to a cavity},\ }\href {https://doi.org/10.1103/PhysRevA.105.023714}
  {\bibfield  {journal} {\bibinfo  {journal} {Phys. Rev. A}\ }\textbf {\bibinfo
  {volume} {105}},\ \bibinfo {pages} {023714} (\bibinfo {year}
  {2022})}\BibitemShut {NoStop}%
\bibitem [{\citenamefont {Clark}\ \emph {et~al.}(2019)\citenamefont {Clark},
  \citenamefont {Jia}, \citenamefont {Schine}, \citenamefont {Baum},
  \citenamefont {Georgakopoulos},\ and\ \citenamefont {Simon}}]{Clark2019}%
  \BibitemOpen
  \bibfield  {author} {\bibinfo {author} {\bibfnamefont {L.~W.}\ \bibnamefont
  {Clark}}, \bibinfo {author} {\bibfnamefont {N.}~\bibnamefont {Jia}}, \bibinfo
  {author} {\bibfnamefont {N.}~\bibnamefont {Schine}}, \bibinfo {author}
  {\bibfnamefont {C.}~\bibnamefont {Baum}}, \bibinfo {author} {\bibfnamefont
  {A.}~\bibnamefont {Georgakopoulos}},\ and\ \bibinfo {author} {\bibfnamefont
  {J.}~\bibnamefont {Simon}},\ }\bibfield  {title} {\bibinfo {title}
  {Interacting {{Floquet}} polaritons},\ }\href
  {https://doi.org/10.1038/s41586-019-1354-5} {\bibfield  {journal} {\bibinfo
  {journal} {Nature}\ }\textbf {\bibinfo {volume} {571}},\ \bibinfo {pages}
  {532} (\bibinfo {year} {2019})}\BibitemShut {NoStop}%
\bibitem [{\citenamefont {Shlesinger}\ \emph {et~al.}(2021)\citenamefont
  {Shlesinger}, \citenamefont {Senellart}, \citenamefont {Senellart},
  \citenamefont {Lanco}, \citenamefont {Lanco}, \citenamefont {Greffet},\ and\
  \citenamefont {Greffet}}]{Shlesinger2021}%
  \BibitemOpen
  \bibfield  {author} {\bibinfo {author} {\bibfnamefont {I.}~\bibnamefont
  {Shlesinger}}, \bibinfo {author} {\bibfnamefont {P.}~\bibnamefont
  {Senellart}}, \bibinfo {author} {\bibfnamefont {P.}~\bibnamefont
  {Senellart}}, \bibinfo {author} {\bibfnamefont {L.}~\bibnamefont {Lanco}},
  \bibinfo {author} {\bibfnamefont {L.}~\bibnamefont {Lanco}}, \bibinfo
  {author} {\bibfnamefont {J.-J.}\ \bibnamefont {Greffet}},\ and\ \bibinfo
  {author} {\bibfnamefont {J.-J.}\ \bibnamefont {Greffet}},\ }\bibfield
  {title} {\bibinfo {title} {Time-frequency encoded single-photon generation
  and broadband single-photon storage with a tunable subradiant state},\ }\href
  {https://doi.org/10.1364/OPTICA.396223} {\bibfield  {journal} {\bibinfo
  {journal} {Optica, OPTICA}\ }\textbf {\bibinfo {volume} {8}},\ \bibinfo
  {pages} {95} (\bibinfo {year} {2021})}\BibitemShut {NoStop}%
\bibitem [{\citenamefont {Pi{\~n}eiro~Orioli}\ \emph
  {et~al.}(2022)\citenamefont {Pi{\~n}eiro~Orioli}, \citenamefont {Thompson},\
  and\ \citenamefont {Rey}}]{PineiroOrioli2022}%
  \BibitemOpen
  \bibfield  {author} {\bibinfo {author} {\bibfnamefont {A.}~\bibnamefont
  {Pi{\~n}eiro~Orioli}}, \bibinfo {author} {\bibfnamefont {J.~K.}\ \bibnamefont
  {Thompson}},\ and\ \bibinfo {author} {\bibfnamefont {A.~M.}\ \bibnamefont
  {Rey}},\ }\bibfield  {title} {\bibinfo {title} {Emergent {{Dark States}} from
  {{Superradiant Dynamics}} in {{Multilevel Atoms}} in a {{Cavity}}},\ }\href
  {https://doi.org/10.1103/PhysRevX.12.011054} {\bibfield  {journal} {\bibinfo
  {journal} {Phys. Rev. X}\ }\textbf {\bibinfo {volume} {12}},\ \bibinfo
  {pages} {011054} (\bibinfo {year} {2022})}\BibitemShut {NoStop}%
\bibitem [{\citenamefont {Lei}\ \emph {et~al.}(2022)\citenamefont {Lei},
  \citenamefont {Fukumori}, \citenamefont {Rochman}, \citenamefont {Zhu},
  \citenamefont {Endres}, \citenamefont {Choi},\ and\ \citenamefont
  {Faraon}}]{Lei2022}%
  \BibitemOpen
  \bibfield  {author} {\bibinfo {author} {\bibfnamefont {M.}~\bibnamefont
  {Lei}}, \bibinfo {author} {\bibfnamefont {R.}~\bibnamefont {Fukumori}},
  \bibinfo {author} {\bibfnamefont {J.}~\bibnamefont {Rochman}}, \bibinfo
  {author} {\bibfnamefont {B.}~\bibnamefont {Zhu}}, \bibinfo {author}
  {\bibfnamefont {M.}~\bibnamefont {Endres}}, \bibinfo {author} {\bibfnamefont
  {J.}~\bibnamefont {Choi}},\ and\ \bibinfo {author} {\bibfnamefont
  {A.}~\bibnamefont {Faraon}},\ }\bibfield  {title} {\bibinfo {title}
  {Many-body cavity quantum electrodynamics with driven inhomogeneous
  emitters},\ }\href@noop {} {\  (\bibinfo {year} {2022})},\ \Eprint
  {https://arxiv.org/abs/2208.04345} {arXiv:2208.04345} \BibitemShut {NoStop}%
\bibitem [{\citenamefont {Barredo}\ \emph {et~al.}(2016)\citenamefont
  {Barredo}, \citenamefont {{de L{\'e}s{\'e}leuc}}, \citenamefont {Lienhard},
  \citenamefont {Lahaye},\ and\ \citenamefont {Browaeys}}]{Barredo2016}%
  \BibitemOpen
  \bibfield  {author} {\bibinfo {author} {\bibfnamefont {D.}~\bibnamefont
  {Barredo}}, \bibinfo {author} {\bibfnamefont {S.}~\bibnamefont {{de
  L{\'e}s{\'e}leuc}}}, \bibinfo {author} {\bibfnamefont {V.}~\bibnamefont
  {Lienhard}}, \bibinfo {author} {\bibfnamefont {T.}~\bibnamefont {Lahaye}},\
  and\ \bibinfo {author} {\bibfnamefont {A.}~\bibnamefont {Browaeys}},\
  }\bibfield  {title} {\bibinfo {title} {An atom-by-atom assembler of
  defect-free arbitrary two-dimensional atomic arrays},\ }\href
  {https://doi.org/10.1126/science.aah3778} {\bibfield  {journal} {\bibinfo
  {journal} {Science}\ }\textbf {\bibinfo {volume} {354}},\ \bibinfo {pages}
  {1021} (\bibinfo {year} {2016})}\BibitemShut {NoStop}%
\bibitem [{\citenamefont {Endres}\ \emph {et~al.}(2016)\citenamefont {Endres},
  \citenamefont {Bernien}, \citenamefont {Keesling}, \citenamefont {Levine},
  \citenamefont {Anschuetz}, \citenamefont {Krajenbrink}, \citenamefont
  {Senko}, \citenamefont {Vuletic}, \citenamefont {Greiner},\ and\
  \citenamefont {Lukin}}]{Endres2016}%
  \BibitemOpen
  \bibfield  {author} {\bibinfo {author} {\bibfnamefont {M.}~\bibnamefont
  {Endres}}, \bibinfo {author} {\bibfnamefont {H.}~\bibnamefont {Bernien}},
  \bibinfo {author} {\bibfnamefont {A.}~\bibnamefont {Keesling}}, \bibinfo
  {author} {\bibfnamefont {H.}~\bibnamefont {Levine}}, \bibinfo {author}
  {\bibfnamefont {E.~R.}\ \bibnamefont {Anschuetz}}, \bibinfo {author}
  {\bibfnamefont {A.}~\bibnamefont {Krajenbrink}}, \bibinfo {author}
  {\bibfnamefont {C.}~\bibnamefont {Senko}}, \bibinfo {author} {\bibfnamefont
  {V.}~\bibnamefont {Vuletic}}, \bibinfo {author} {\bibfnamefont
  {M.}~\bibnamefont {Greiner}},\ and\ \bibinfo {author} {\bibfnamefont {M.~D.}\
  \bibnamefont {Lukin}},\ }\bibfield  {title} {\bibinfo {title} {Atom-by-atom
  assembly of defect-free one-dimensional cold atom arrays},\ }\href
  {https://doi.org/10.1126/science.aah3752} {\bibfield  {journal} {\bibinfo
  {journal} {Science}\ }\textbf {\bibinfo {volume} {354}},\ \bibinfo {pages}
  {1024} (\bibinfo {year} {2016})}\BibitemShut {NoStop}%
\bibitem [{\citenamefont {Deist}\ \emph {et~al.}(2022)\citenamefont {Deist},
  \citenamefont {Gerber}, \citenamefont {Lu}, \citenamefont {Zeiher},\ and\
  \citenamefont {{Stamper-Kurn}}}]{Deist2022}%
  \BibitemOpen
  \bibfield  {author} {\bibinfo {author} {\bibfnamefont {E.}~\bibnamefont
  {Deist}}, \bibinfo {author} {\bibfnamefont {J.~A.}\ \bibnamefont {Gerber}},
  \bibinfo {author} {\bibfnamefont {Y.-H.}\ \bibnamefont {Lu}}, \bibinfo
  {author} {\bibfnamefont {J.}~\bibnamefont {Zeiher}},\ and\ \bibinfo {author}
  {\bibfnamefont {D.~M.}\ \bibnamefont {{Stamper-Kurn}}},\ }\bibfield  {title}
  {\bibinfo {title} {Superresolution {{Microscopy}} of {{Optical Fields Using
  Tweezer-Trapped Single Atoms}}},\ }\href
  {https://doi.org/10.1103/PhysRevLett.128.083201} {\bibfield  {journal}
  {\bibinfo  {journal} {Phys. Rev. Lett.}\ }\textbf {\bibinfo {volume} {128}},\
  \bibinfo {pages} {083201} (\bibinfo {year} {2022})}\BibitemShut {NoStop}%
\bibitem [{\citenamefont {Sauerwein}\ \emph {et~al.}(2022)\citenamefont
  {Sauerwein}, \citenamefont {Orsi}, \citenamefont {Uhrich}, \citenamefont
  {Bandyopadhyay}, \citenamefont {Mattiotti}, \citenamefont
  {{Cantat-Moltrecht}}, \citenamefont {Pupillo}, \citenamefont {Hauke},\ and\
  \citenamefont {Brantut}}]{Sauerwein2022}%
  \BibitemOpen
  \bibfield  {author} {\bibinfo {author} {\bibfnamefont {N.}~\bibnamefont
  {Sauerwein}}, \bibinfo {author} {\bibfnamefont {F.}~\bibnamefont {Orsi}},
  \bibinfo {author} {\bibfnamefont {P.}~\bibnamefont {Uhrich}}, \bibinfo
  {author} {\bibfnamefont {S.}~\bibnamefont {Bandyopadhyay}}, \bibinfo {author}
  {\bibfnamefont {F.}~\bibnamefont {Mattiotti}}, \bibinfo {author}
  {\bibfnamefont {T.}~\bibnamefont {{Cantat-Moltrecht}}}, \bibinfo {author}
  {\bibfnamefont {G.}~\bibnamefont {Pupillo}}, \bibinfo {author} {\bibfnamefont
  {P.}~\bibnamefont {Hauke}},\ and\ \bibinfo {author} {\bibfnamefont {J.-P.}\
  \bibnamefont {Brantut}},\ }\bibfield  {title} {\bibinfo {title} {Engineering
  random spin models with atoms in a high-finesse cavity},\ }\href@noop {} {\
  (\bibinfo {year} {2022})},\ \Eprint {https://arxiv.org/abs/2208.09421}
  {arXiv:2208.09421} \BibitemShut {NoStop}%
\bibitem [{\citenamefont {Ferri}\ \emph {et~al.}(2022)\citenamefont {Ferri},
  \citenamefont {Rooij}, \citenamefont {Lebouteiller}, \citenamefont {Bourdel},
  \citenamefont {Baghdad}, \citenamefont {Schwartz}, \citenamefont {Garcia},
  \citenamefont {Reichel},\ and\ \citenamefont {Long}}]{ferri2022}%
  \BibitemOpen
  \bibfield  {author} {\bibinfo {author} {\bibfnamefont {F.}~\bibnamefont
  {Ferri}}, \bibinfo {author} {\bibfnamefont {A.~L.}\ \bibnamefont {Rooij}},
  \bibinfo {author} {\bibfnamefont {C.}~\bibnamefont {Lebouteiller}}, \bibinfo
  {author} {\bibfnamefont {P.-A.}\ \bibnamefont {Bourdel}}, \bibinfo {author}
  {\bibfnamefont {M.}~\bibnamefont {Baghdad}}, \bibinfo {author} {\bibfnamefont
  {S.}~\bibnamefont {Schwartz}}, \bibinfo {author} {\bibfnamefont
  {S.}~\bibnamefont {Garcia}}, \bibinfo {author} {\bibfnamefont
  {J.}~\bibnamefont {Reichel}},\ and\ \bibinfo {author} {\bibfnamefont
  {R.}~\bibnamefont {Long}},\ }\bibfield  {title} {\bibinfo {title} {An optical
  elevator for precise delivery of cold atoms using an acousto-optical
  deflector},\ }\href {https://doi.org/10.1088/1367-2630/ac5f84} {\bibfield
  {journal} {\bibinfo  {journal} {New J. Phys.}\ }\textbf {\bibinfo {volume}
  {24}},\ \bibinfo {pages} {043013} (\bibinfo {year} {2022})}\BibitemShut
  {NoStop}%
\bibitem [{\citenamefont {Johansson}\ \emph {et~al.}(2013)\citenamefont
  {Johansson}, \citenamefont {Nation},\ and\ \citenamefont
  {Nori}}]{Johansson2013}%
  \BibitemOpen
  \bibfield  {author} {\bibinfo {author} {\bibfnamefont {J.~R.}\ \bibnamefont
  {Johansson}}, \bibinfo {author} {\bibfnamefont {P.~D.}\ \bibnamefont
  {Nation}},\ and\ \bibinfo {author} {\bibfnamefont {F.}~\bibnamefont {Nori}},\
  }\bibfield  {title} {\bibinfo {title} {{{QuTiP}} 2: {{A Python}} framework
  for the dynamics of open quantum systems},\ }\href
  {https://doi.org/10.1016/j.cpc.2012.11.019} {\bibfield  {journal} {\bibinfo
  {journal} {Computer Physics Communications}\ }\textbf {\bibinfo {volume}
  {184}},\ \bibinfo {pages} {1234} (\bibinfo {year} {2013})}\BibitemShut
  {NoStop}%
\bibitem [{\citenamefont {Arora}\ \emph {et~al.}(2007)\citenamefont {Arora},
  \citenamefont {Safronova},\ and\ \citenamefont {Clark}}]{Arora2007}%
  \BibitemOpen
  \bibfield  {author} {\bibinfo {author} {\bibfnamefont {B.}~\bibnamefont
  {Arora}}, \bibinfo {author} {\bibfnamefont {M.~S.}\ \bibnamefont
  {Safronova}},\ and\ \bibinfo {author} {\bibfnamefont {C.~W.}\ \bibnamefont
  {Clark}},\ }\bibfield  {title} {\bibinfo {title} {Magic wavelengths for the
  \$np\textbackslash text\{\textbackslash ensuremath\{-\}\}ns\$ transitions in
  alkali-metal atoms},\ }\href {https://doi.org/10.1103/PhysRevA.76.052509}
  {\bibfield  {journal} {\bibinfo  {journal} {Phys. Rev. A}\ }\textbf {\bibinfo
  {volume} {76}},\ \bibinfo {pages} {052509} (\bibinfo {year}
  {2007})}\BibitemShut {NoStop}%
\bibitem [{\citenamefont {Le~Kien}\ \emph {et~al.}(2013)\citenamefont
  {Le~Kien}, \citenamefont {Schneeweiss},\ and\ \citenamefont
  {Rauschenbeutel}}]{LeKien2013}%
  \BibitemOpen
  \bibfield  {author} {\bibinfo {author} {\bibfnamefont {F.}~\bibnamefont
  {Le~Kien}}, \bibinfo {author} {\bibfnamefont {P.}~\bibnamefont
  {Schneeweiss}},\ and\ \bibinfo {author} {\bibfnamefont {A.}~\bibnamefont
  {Rauschenbeutel}},\ }\bibfield  {title} {\bibinfo {title} {Dynamical
  polarizability of atoms in arbitrary light fields: General theory and
  application to cesium},\ }\href {https://doi.org/10.1140/epjd/e2013-30729-x}
  {\bibfield  {journal} {\bibinfo  {journal} {Eur. Phys. J. D}\ }\textbf
  {\bibinfo {volume} {67}},\ \bibinfo {pages} {92} (\bibinfo {year}
  {2013})}\BibitemShut {NoStop}%
\bibitem [{\citenamefont {Gardiner}\ and\ \citenamefont
  {Zoller}(2004)}]{Gardiner2004}%
  \BibitemOpen
  \bibfield  {author} {\bibinfo {author} {\bibfnamefont {C.}~\bibnamefont
  {Gardiner}}\ and\ \bibinfo {author} {\bibfnamefont {P.}~\bibnamefont
  {Zoller}},\ }\href@noop {} {\emph {\bibinfo {title} {Quantum {{Noise}}: {{A
  Handbook}} of {{Markovian}} and {{Non-Markovian Quantum Stochastic Methods}}
  with {{Applications}} to {{Quantum Optics}}}}},\ \bibinfo {edition} {3rd}\
  ed.,\ Springer {{Series}} in {{Synergetics}}\ (\bibinfo  {publisher}
  {{Springer-Verlag}},\ \bibinfo {address} {{Berlin Heidelberg}},\ \bibinfo
  {year} {2004})\BibitemShut {NoStop}%
\bibitem [{\citenamefont {Law}\ \emph {et~al.}(1995)\citenamefont {Law},
  \citenamefont {Zhu},\ and\ \citenamefont {Zubairy}}]{Law1995}%
  \BibitemOpen
  \bibfield  {author} {\bibinfo {author} {\bibfnamefont {C.~K.}\ \bibnamefont
  {Law}}, \bibinfo {author} {\bibfnamefont {S.-Y.}\ \bibnamefont {Zhu}},\ and\
  \bibinfo {author} {\bibfnamefont {M.~S.}\ \bibnamefont {Zubairy}},\
  }\bibfield  {title} {\bibinfo {title} {Modification of a vacuum {{Rabi}}
  splitting via a frequency-modulated cavity mode},\ }\href
  {https://doi.org/10.1103/PhysRevA.52.4095} {\bibfield  {journal} {\bibinfo
  {journal} {Phys. Rev. A}\ }\textbf {\bibinfo {volume} {52}},\ \bibinfo
  {pages} {4095} (\bibinfo {year} {1995})}\BibitemShut {NoStop}%
\end{thebibliography}%

  \vspace{5cm}

\onecolumngrid 
\newpage

%-----------------------------------------------------------------------
%---------------------------------------------------------------------
%         METHODS
%---------------------------------------------------------------------
%---------------------------------------------------------------------

\begin{center}
    \vspace{0mm}
    %\hrule \vspace{5mm}
    \large{\textbf{METHODS}}
    \vspace{3mm}\
\end{center}

\twocolumngrid

\subsection*{Experimental setup}
\vspace{-2mm}

At the core of our experimental setup lies a high-finesse fiber Fabry-Perot optical microcavity with a length of $140\,\mu$m.  The maximum single atom coupling strength is $g_0/2\pi\simeq 76\,$MHz, larger than the cavity decay rate $\kappa/2\pi \simeq 15\,$MHz and the spontaneous emission rate $\gamma/2\pi \simeq 3\,$MHz ($\kappa$ and $\gamma$ are HWHM values), so that we operate in the strong coupling regime of CQED at the single atom level. To realize a homogeneous and maximal coupling of all atoms with the cavity mode, the atoms are trapped at the antinodes of the cavity mode which is resonant with the D2 line of $^{87}$Rb at 780$\,$nm (see Fig. 1 in the main text). We achieve this condition by using an intracavity standing wave at a wavelength of 1559\,nm, commensurate with the atomic line taking into account the Gouy phase shift and by optimizing the phase shift at reflection on the mirrors \cite{Garcia2020a, Ferri2020a} such that the antinodes of both 1559\,nm and 780\,nm standing waves overlap optimally.  

To load the atoms into the cavity, we start with a 3D magneto-optical trap (MOT) of Rubidium atoms positioned 12$\,$mm below the cavity and loaded by a 2D-MOT. The atoms are then trapped inside a horizontal dipole beam that we can displace vertically with an acousto-optical deflector \cite{ferri2022}. In 100$\,$ms, the atoms are moved into the cavity. To load the atoms into the one dimensional intracavity lattice, we increase the 1559$\,$nm intracavity power while decreasing the intensity of the transport beams. We can adjust the number of atoms from a few tens up to 2000 by varying the loading parameters of the MOT. The atoms are first trapped  in a 300$\, \mu K$ deep intracavity lattice at a temperature of 50$\, \mu K$. We can then increase the trap depth by performing an adiabatic compression of the cloud.

\subsection*{Atomic frequency distributions probed by loss spectroscopy}
\vspace{-3mm}

The strong frequency broadening in our experiment results from the combination of finite atomic temperature, lightshifts and level mixing effects described in the Supplementary Information. To characterize the frequency distribution of the atoms, we measure the losses of the atom ensemble (prepared in the $F=2$ hyperfine ground state) induced by a probe beam perpendicular to the cavity axis. This beam is switched on for $0.5\,$ms at a given frequency before performing the loss measurement. We repeat this process for frequencies values across the distribution. To obtain reliable atom number measurements down to low atom numbers, the loss measurement is performed by measuring the vacuum Rabi splitting of the cavity transmission signal (see section Cavity Transmission Spectrum). The results are shown on Fig. \ref{fig:Figure1}. Because the atoms are excited with uncontrolled phases and amplitudes by this transverse beam, we expect (and observe) negligible coupling between the latter and the bright modes of the coupled atom-cavity system even though the resonance condition is fulfilled. 

To generate the simulated frequency distribution curves shown on Fig. \ref{fig:Figure1}c, we assume that the atoms are in thermal equilibrium at each site of the optical lattice, described by an harmonic trap with radial frequencies $\omega_x=\omega_y$ and longitudinal frequency $\omega_z$. For a trap depth of $1.40\,$mK, $\omega_{x,y}/2\pi=14.5\,$kHz, and $ \omega_{z}/2\pi=330\,$ kHz. For each atom of a given sample, we first draw $x$, $y$ and $z$ from a normal distribution with variance 
$\sigma_{x,y,z}^2={k_B T/(m\omega_{x,y,z}^2)}$. Then we deduce the values of the coupling $g(x,y,z)$ (based on the cavity parameters) and of the dipole trap intensity $I_{dip}(x,y,z)$ (based on the maximum value at the bottom of the trap, which is estimated from lightshift measurements and corroborated by a direct transmission measurement). The intensity $I_{dip}(x,y,z)$ seen by each atom is used to construct a $16\times16$ matrix representing the Stark operator in the $\vert F,m_F\rangle$ basis of the $5P_{3/2}$ manifold (see Supplementary Information). The output of the procedure for $N$ atoms is a collection of $16\times N$ eigenvalues  $\hbar \omega_{k,j}$ and eigenvectors $\vert \psi_{k,j}\rangle$, where $1\leq j \leq16$ and $1\leq k\leq N$.

In addition, we assume that the initial atomic population is equally distributed between all Zeeman sublevels of the $F=2$ hyperfine ground state. For each atom (labelled by the index $k$), we use the previous procedure to compute the $5 \times 16$ transition frequencies $\omega_{k,j}(m_F)$ between the 5 ground states $\vert F=2,m_F\rangle$ and the 16 excited states $\vert \psi_{k,j}\rangle$ (with $-2\leq m_F \leq 2$ and $1 \leq j \leq 16$). Each frequency $\omega_{k,j}(m_F)$ is associated to a coupling strength $c_{k,j}(m_F)$, defined as:
\begin{equation} \label{eq_def_c}
c_{k,j}(m_F)= \left \vert  \left \langle \psi_{k,j}  \left \vert \sum_{q=-1}^1\hat{d}_q E_q^t  \right \vert F=2,m_F  \right \rangle \right \vert ^2 \;,
\end{equation} 
where $E_q^t$ are the components of the electric field of the probe beam $\mathbf E^t$, expressed in the spherical basis and $\hat{d}_q$ are the dipole matrix elements (see Supplementary Material). In the experiment,  we have $E_0^t/ \vert \mathbf E^t \vert =1/\sqrt{2}$ and $E_{\pm 1}^t/ \vert \mathbf E^t \vert =\pm 1/2$. The frequency distributions shown on Fig. \ref{fig:Figure1}.c are then obtained by sorting the transition frequencies $\omega_{k,j}(m_F)$ in equal-width bins, weighted by their coupling strength $c_{k,j}(m_F)$. For a given trap depth, we use an effective temperature $T$ (2 to 3 times the initial experimental temperature) to  match the positions of the simulated and experimental curves. The discrepancy stems from heating and depumping effects induced by the transverse beam. The amplitude of each curve is adjusted such that they all have the same area.

\subsection*{Cavity Transmission Spectrum}\label{sec:spectrum}

\subsubsection*{Data acquisition}

To measure the cavity transmission spectrum, we first prepare the atoms in the $\vert F=2,m_F=2\rangle$ Zeeman sublevel by optical pumping. The trap depth is $1400\pm 30 \,\mu$K and the temperature  $190\pm 20\,\mu$K. To ensure that no atom remains in other sublevels, we furthermore apply the following sequence: microwave tranfer to $\vert F=1,m_F=1\rangle$, blast of the remaining atoms in the $\vert F=2\rangle$ hyperfine level and microwave transfer back to $\vert F=2,m_F=2\rangle$. We estimate that $\approx 800$  atoms are in the $\vert F=2,m_F=2\rangle$ state at the end of the preparation sequence. We then probe the cavity-atoms coupled system by sweeping a tunable laser over a range of $8\,$GHz in $8\,$ms. Its intensity is chosen such that the average photon number in the cavity is smaller than unity ($\langle n_{cav} \rangle = $ 0.14), to be in low excitation limit. We measure the transmitted light with a single-photon counting module. To calibrate the frequency axis of the spectrum, we record simultaneously the transmission signal of a 10$\,$cm long cavity providing a frequency ruler allowing us to compensate the slight non-linearity of the frequency sweep, and a saturated absorption signal of the probe laser that provides an absolute frequency reference. We estimate the uncertainty of the frequency of the probe laser to be $\pm$8 MHz (1 standard deviation), well below the Rabi splittings of typically 100 to 4000$\,$MHz. 

As the single-shot spectra are strongly discretized (see Fig. \ref{fig:single_shot}), we average $\approx $  200 such spectra to obtain the experimental data shown on Fig. \ref{fig:Figure2}. As experimental fluctuations of $N$ result into fluctuations of the collective coupling, we group and average the spectra according to their collective coupling $\Omega$ to avoid excessive broadening due to averaging. We used a $40\,$MHz bin centerered on $\Omega/2\pi = 1670\,$MHz, to maximize the number of spectra in the bin. We fit independently the high and low frequency peaks of the averaged spectrum with a Voigt profile to extract the half-width at half-maximum (HWHM) of the peaks. The HWHM width ($\delta \omega_- /2\pi$ = $28\pm2\,$MHz) of the low frequency peaks is a bit larger than the one of the high frequency peak ($\delta \omega_+ /2\pi$ = $24\pm2\,$MHz), probably due to the presence of atom losses during the measurement. The HWHM value quoted in the main text ($\delta \omega/2\pi$ = $26\pm2\,$MHz) is the average of the experimental polaritons' widths.

\subsubsection*{Simulation of the transmission spectrum and frequency distribution}

For a given atom $k$, the excited subspace of the $5P_{3/2}$ manifold dressed by the dipole light features 16 eigenvalues $\hbar \omega_{k,j}$  and the 16 corresponding eigenstates $\vert \psi_{k,j}\rangle$, where $1\leq k \leq N$ enumerates the atoms and  $1\leq j \leq 16$ the eigenstates of the excited subspace. The cavity is probed with a weak pulse of $\sigma_+$ polarized light which couples the ground state $\vert k:2,2\rangle$ of atom number $k$ to $\vert k:3,3\rangle$, the $\vert 5P_{3/2},F'=3,m_{F'}=3\rangle$ Zeeman sublevel of this atom. The coherent evolution is described by a multilevel version of the standard Tavis-Cummings Hamiltonian (see Supplementary Information). Taking into account photon losses $\kappa$ and atomic decay $\gamma$, the transmission of the cavity as a function of the probe frequency $\omega$, is given by the following simple analytical form, that was derived in \cite{Diniz2011b} using the input-output formalism:
\begin{equation} \label{eq:transmission}
t(\omega)=\frac{-\kappa/(2i)}{ \displaystyle \omega_c-i\kappa/2-\omega-\sum_{k,j}\frac{\left(g_{k,j}\right)^2}{\omega_{k,j}-i\gamma/2-\omega}} .
\end{equation}
To simulate the averaged spectrum of Fig. \ref{fig:Figure2}, we compute many transmission spectra with a trap depth $U_0 = 1400\,\mu$K and a temperature $T = 190\,\mu$K. We draw randomly the number of atoms $N$, to account for experimental fluctuations. We average single spectra ($\approx 350$) featuring a collective coupling within the bin $1670\pm28\,$MHz. This corresponds to the bin chosen for the experimental spectra of Fig. 2, enlarged by $\pm8\,$MHz to account for the 1-standard deviation uncertainty on frequencies of the spectrum. 
For the spectral distribution, we reconstruct the spectral distribution of couplings $\rho(\omega)=\sum_{k=1}^{N} \sum_{j=1}^{16} g_{k,j}^{2} \delta(\omega-\omega_{k,j})$ by drawing $N = 100\,000$ random atomic positions and computing the associated distribution of $(\omega_{k,j},g_{k,j})$. We then sort the transition frequencies $\omega_{k,j}$ in equal-width bins, weighted by their coupling strength $\vert g_{k,j}\vert ^2$. The calculated distributions are shown on Fig. \ref{fig:atomic_distribution} for various values of the trap depth $U_0$, and on Fig. \ref{fig:Figure2} ($U_0 = 1040\,\mu K$) together with the corresponding transmission spectrum.

\subsection*{Transition from the unprotected to the cavity-protected regime for a finite-sized system}

\subsubsection*{Data acquisition}

To explore the transition from the unprotected to the cavity-protected regime, we acquire $\sim 2000$ experimental spectra for an increasing number of atoms $N$ (from $\sim10$ to 550). Experimentally we tune $N$ by varying the loading time of our magneto-optical-trap. The trap depth is $U_0 = 1040\pm 30\, \mu$K and the temperature is $T = 140\pm 20\,\mu$K. The cavity frequency is set equal to the average frequency of the corresponding coupling distribution $\rho(\omega)$ (red distribution in Fig 1.): $\omega_c = \omega_{\vert 2,2\rangle \rightarrow \vert 3,3\rangle} - 970\,$MHz (we justify this choice for $\omega_c$ in Supplemental Information section II.2).  For each experimental  spectrum, we compute the collective coupling $\Omega$ as the distance between the central frequencies $\omega_{\pm}$ of each half of the spectrum. $\omega_{\pm}$ is defined as the barycenter of the photon-count distribution within a half-spectrum. This definition does not rely on a fit and allows computing $\Omega$ even for multi-peak spectra in the unprotected regime. 

As the average number of photons inside the cavity is much lower than one, the number of photons collected is low and the spectrum is strongly discretized as shown in Fig.\ref{fig:single_shot}, preventing us from directly extracting the $PW$ of eigenstates exhibiting a low transmission. 
So, we characterize the degree of coherence of the system by measuring the fraction $F_{out}$ of photon counts outside a frequency range $\Delta f/ 2\pi$, in which most of the counts lay in the protected regime. First, we define the typical frequency width $\Delta f/2\pi$ of the polariton peak in the cavity-protected regime (for $\Omega/2\pi \geq 1000\,$MHz) as twice the standard deviation $\sigma$ of the photon-count weighted frequency distribution. In this regime, we measure $\Delta f/2\pi = 140\,$MHz and $\approx 90 \%$ of the counts are within $\Delta f$. Then we separate the spectrum into two halves at the cavity frequency $\omega_c$. For each half of the spectrum ($\pm$), we compute the total number of counts $N_{tot, \pm}$, the mean value $\Omega_{\pm}$ of the frequency distribution of the spectrum (weighted by the counts) and the number of counts $N_{out, \pm}$ that are outside the intervals $[\Omega_{\pm} - \Delta f/2 ;\Omega_{\pm} + \Delta f/2]$. Finally, we compute the fractions of counts outside of these intervals, $F_{out, \pm}$ (respectively $F_{out}$), for the two halves of the spectrum (respectively the total spectrum). For $F_{out}$, we get a cloud of points $(\Omega, F_{out})$, which we average with $30\,$MHz-bins of $\Omega$. Vertical error bars are equal to $\pm\frac{\sigma}{\sqrt{N_{p}}}$, where $\sigma$ is the standard deviation of the  $N_{p}$ points gathered within a bin. The resulting curve is plotted in the Fig. \ref{fig:Figure3}. We have also checked that $F_{out}$ is robust with respect to the exact value of $\Delta f/ 2\pi$, as shown in Fig.~\ref{fig:comparing_exclusion_window}).

\subsubsection*{Calculation of  $S_{PW}$}

Operating in the strong coupling regime at the single atom level enables us to explore the transition from the unprotected to the protected regime for a mesoscopic number of atoms ranging from 10 to 550. For such numbers of emitters, we are far from the continuous frequency distribution of emitters usually encountered in solid-state systems  and are sensitive to the finite size of the Hilbert space and to the discreteness of the spectrum of the coupled emitters-cavity system. In particular, we can directly compute the  eigenvalues and eigenvectors of the coupled system and extract their properties.

In the low excitation limit, we restrict the dynamics of the closed system to the one excitation manifold. For $N$ two-level atoms, it is spanned by the $N+1$ basis states  $\left\{ \vert 1, G \rangle, \vert 0, E_{1}\rangle, \ldots, \vert 0, E_{k}\rangle, \ldots, \vert 0, E_{N}\rangle \right\}$ with $\vert G\rangle=\vert g_{1} \ldots g_{N} \rangle$ and $\vert E_{k}\rangle = \vert g_{1} \ldots e_{k} \ldots g_{N}\rangle$, where $\ket{g_k}$ (resp. $\ket{e_k}$) is the ground (resp. excited) state of the atom $k$.

Due to the 1559$\,$nm trapping light, a given atom $k$  is excited to the subspace of the $5P_{3/2}$ manifold featuring 16 eigenvalues $\omega_{k,j}$ and eigenstates $\vert \psi_{k,j}\rangle$. The single excitation manifold is then spanned by the $16 \times N +1$ basis states $ \left\{ \vert 1,  G \rangle, \vert 0, E_{1,1} \rangle, \ldots, \vert 0, E_{k,j} \rangle \ldots  \vert 0, E_{N,16} \rangle  \right\}$  where $\vert G \rangle= \vert G_{1}, \ldots, G_{k}, \ldots,  G_{N} \rangle$ with $\vert G_{k}\rangle = \vert  g_{k,1} \ldots g_{k,j}  \ldots  g_{k,16} \rangle $ and
$ \vert E_{k,j} \rangle = \vert G_{1}, \ldots, g_{k,1} \ldots \psi_{k,j} \ldots g_{k,16}, \ldots, G_{N} \rangle$. 

In this basis, the Tavis-Cummings Hamiltonian for mutilevels atoms is given by an arrowhead matrix, from which we can numerically extract the $D=16 \times N + 1$ eigenstates $\Psi_i^{C}$  and eigenvalues $\varepsilon_i$ ($i=1,\ldots,D$) of the atoms - cavity coupled system. We can then compute the photonic weight of the eigenstates  $\Psi_i^{C}$  defined as the weight of the $\vert 1, G \rangle $ component: 
\[ PW_i=\vert \langle 1,G \vert \Psi_i^{C}  \rangle \vert^2 \]

To simulate the theoretical curve of Fig. \ref{fig:Figure3}, we draw $N_{atom}$ random atomic positions (where $N_{atom}$ is the atom number), deduce the associated values of frequencies and cavity couplings $(\omega_{k,j},g_{k,j})$ and compute the $PW$ of the different eigenstates. We then  calculate $S_{PW}$ which we have defined as the sum of all photonic weights except the two largest ones: $S_{PW}=1-PW_{max1}-PW_{max2}$. For each value of $N_{atom}$, we average $S_{PW}$ over 300 repetitions.

\subsection*{Polariton Modulation}

\subsubsection*{Data acquisition}

To modulate the polaritons, we modulate the average light-shifted frequency of the atoms $\overline{\omega_A}(t)=\overline{\omega_A^0}+ \beta_o \omega_m \cos(\omega_m th)$ where $\beta_o$ is the modulation index, by modulating the intracavity dipole power at a frequency $\omega_m/2 \pi$. The spectrum of Fig. \ref{fig:Figure4} is obtained by averaging $\approx$ 200 spectra. Before averaging, we select spectra with similar atom number $N$, by measuring the dispersive shift $\delta\omega_{c}$ of the cavity frequency when the atoms are in the state $\vert F=1, m_F=1\rangle$, which is given by: 

\begin{eqnarray}
\delta\omega_{c} =\frac{-N g_{\vert2,2\rangle \rightarrow \vert 3,3\rangle}^2}{2(\omega_{\vert1,1\rangle \rightarrow \vert 2,2\rangle} - \omega_c) }
\end{eqnarray}

We select spectra within the dispersive shift bin $ \delta\omega_{c}/2\pi = -258\pm 5\,$MHz, corresponding to $N=1130\pm25$ or to collective coupling $ \Omega/2\pi = 2020\pm20\,$MHz. The trap depth is 1400$\,\mu$K and the temperature is 190$\,\mu$K, similar to the data of main text Fig. 2. The modulation frequency is $\omega_{m}/ 2 \pi = 120$ MHz.

As we are in the regime where $\Omega \gg \omega_m $, the cavity spectrum is given by the following expression, derived in Supplementary Information (equation 25.):
\begin{eqnarray}
S(\omega)  \propto   \sum_n   
 \frac{ J_n^2   \left(\beta_o/2 \right)}{(\omega-\omega_0-n\omega_m+ \Omega )^2 + \gamma_s^2} \\ \nonumber
+ \sum_n  \frac{ J_n^2   \left(\beta_o/2 \right)}{(\omega-\omega_0-n\omega_m - \Omega)^2+ \gamma_s^2} 
\label{cavity_spectrum_exper_inter_method}
\end{eqnarray}

The cavity spectrum features two combs of peaks centered at $\pm \Omega$. The separation between two consecutive peaks is given by the modulation frequency $\omega_m$. The amplitude of each peak is given by $J_n^2  \left(\beta_o/2 \right) $ and so the modulation index for the polaritons is given by $\beta_p=\beta_o/2$. We are then able to directly control the modulation of the polaritons in frequency and amplitude with  the modulation of the trapping light. The width of each peak in given by $\gamma_s=\frac{\kappa+\gamma}{2}$, identical to the width of the non-modulated homogeneous case as we are in the cavity protected regime. As the number of atoms inside the cavity modes fluctuates, the transmission peak of the high frequency polariton is best-fitted by Voigt function: 
\begin{eqnarray}
\sum_n  J_n^2   \left(\beta_p \right) \mathrm{Voigt}(\omega-\omega_0-n\omega_m \pm \Omega, \gamma_s, \sigma)
\end{eqnarray}\label{cavity_spectrum_exper_last}
where $\mathrm{Voigt}((\omega, \gamma_s, \sigma)= \frac{1}{\sqrt{2 \pi}\sigma}\Re(\mathrm{wofz}(\frac{\omega+ i\gamma_s}{\sqrt{2} \sigma}))$ with $\mathrm{wofz}$ the Faddeeva function. This function provides a very good fit of the experimental data, except for the slight asymmetric shape (relative variation of $\pm 7\,$\%) between the right and left part of each comb.

\subsubsection*{Simulation of the cavity transmission spectra}

To justify the function used to fit the experimental data and to confirm the analysis presented in the Supplementary Information, we simulate the cavity spectrum by numerically integrating the master equation using QutiP \cite{Johansson2013}, for an emitter-cavity system probed by a laser swept in frequency at the same rate as in the experiment ($1\,$GHz/ms), and with the same power. We calculate the population of the state $\vert 1, G \rangle$ after binning the simulated results to match the experimental resolution of $2.5\,$MHz. By using the parameters given by the experimental fit, we obtain a very good agreement between the experimental data and the simulated spectra even for the asymmetrical shape of the comb.

\subsubsection*{Modulation transfer measurement}

To measure the linear modulation transfer $\beta_p \propto \beta_0$ (inset of Fig. \ref{fig:Figure4}), we measure $\approx$ 100 spectra for each value of $\beta_0$. We reduce the frequency range of the probe laser scan by a factor of 4 to increase the frequency resolution. Before averaging spectra, we frequency-shift each spectrum so that the multipeaks centers of all spectra are aligned, to compensate for atom number fluctuations. Compared to a raw averaging, this technique improves the accuracy of the relative amplitudes of the different peaks in the average spectrum, to which the value of $\beta_{p}$ is very sensitive. We then fit the cavity spectrum with the function (5) to extract the value of $\beta_p$.

Since $\beta_p$ is the result of a non-linear fit, we resort to a nonparametric bootstrap method to determine its uncertainty. For each averaged spectrum (and thus each value of $\beta_p$), we generate $500$ synthetic spectra $A_{k}$  $(1\leq k\leq 500)$. Then, each synthetic spectrum $A_{k}$ is fitted with the formula (5), providing a fitted parameter $\beta_{p,k}$. The errorbar for $\beta_p$ is defined as the $\pm$ the standard deviation of the set $\{\beta_{p,k} $ ; $ 1\leq k \leq 500\}$. To obtain the value of $\beta_{0}$, we measure the intensities of the different frequency components of the trapping light, using the transmission spectrum of the frequency-scanned cavity. Starting from
$ \overline{ \delta \omega_{a} (t)} = \overline{\omega_A}(t) - \overline{\omega_A^0} =\beta_{0} \omega_{m}$cos($\omega_{m}t$), we get:
\begin{eqnarray}
\beta_{0} = \frac{\delta \omega_{a}^{max}}{\omega_{m}} = \frac{\vert \overline{\omega_A^0}\vert  }{\omega_{m}}  \frac{\delta \omega_{a}^{max}}{\vert \overline{\omega_A^0} \vert}          
= \frac{\vert \overline{\omega_A^0} \vert}{\omega_{m}} \frac{\delta I_D^{max}}{I_D^{0}}
\end{eqnarray}
where $\overline{\omega_A^0} / 2 \pi  = -1320\,$MHz is the average atomic lightshift, $\omega_{m}/ 2 \pi = 130$ MHz is the modulation frequency, and $I_{D}^{0}$ and $\delta I_D^{max}$ are, respectively, the constant and $\omega_m$  frequency-dependent components of the intensity of the trapping light. Uncertainties on $\beta_{0}$ are propagated from the experimental uncertainties of $\omega_{a}^{0}$, $I_{D}^{0}$ and $\delta I_D^{max}$.

\onecolumngrid 
\newpage

%-----------------------------------------------------------------------
%---------------------------------------------------------------------
%         EXTENDED DATA
%---------------------------------------------------------------------
%---------------------------------------------------------------------

\begin{center}
    \vspace{0mm}
    %\hrule \vspace{5mm}
    \large{\textbf{EXTENDED DATA}}
    \vspace{3mm}
\end{center}

\setcounter{figure}{0}

\renewcommand{\thefigure}{E\arabic{figure}}

\twocolumngrid 

\begin{figure}[h]
\centering
\includegraphics[width=\columnwidth]{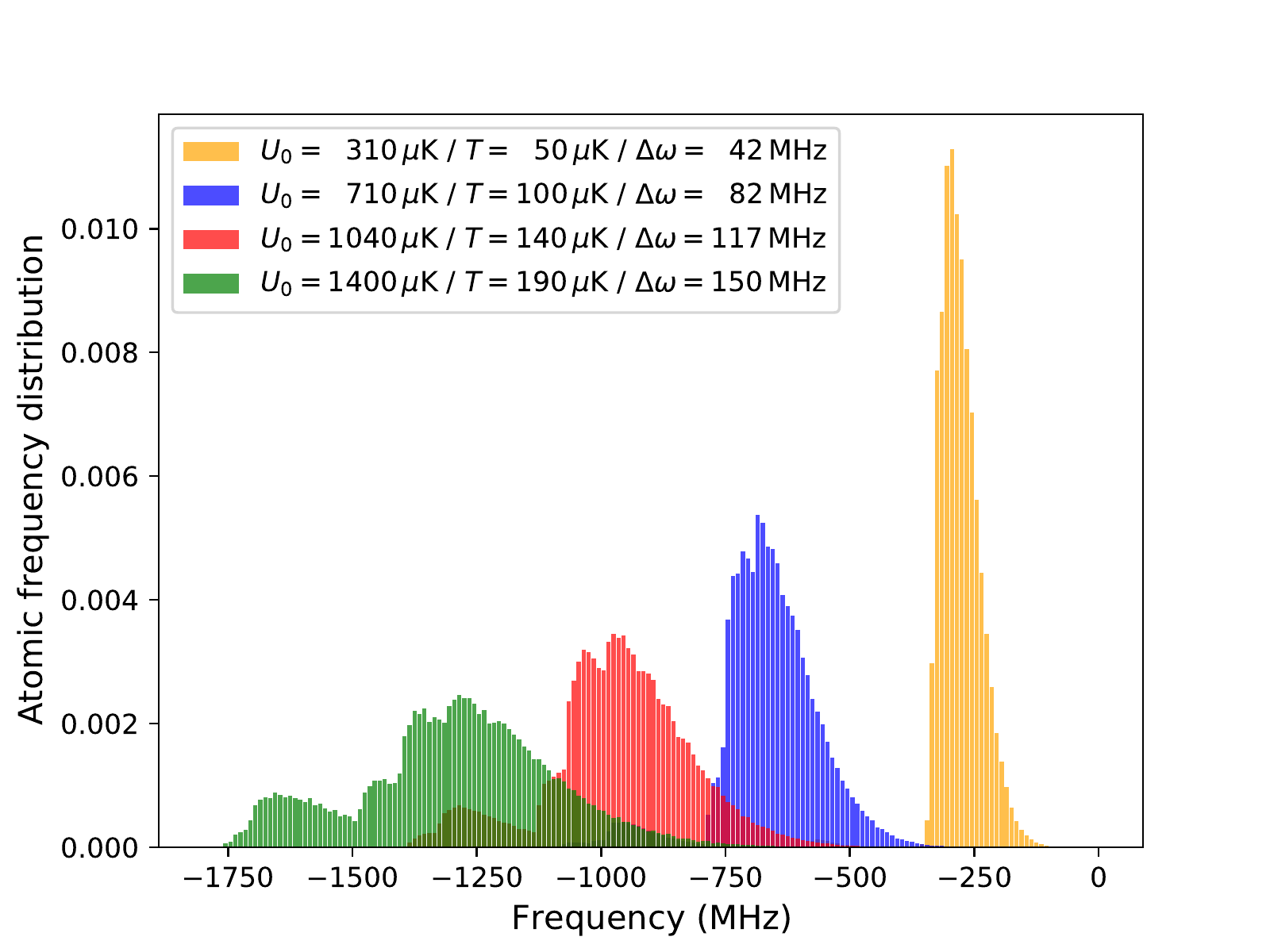}
\caption{\textbf{Simulated atomic frequency distribution for different trap depths}. For each trap depth $U_0$, the temperature $T$ used in the simulation corresponds to the typical experimental value based on time-of-flight measurements. When $U_0$ increases, the mean frequency of the distribution decreases linearly -- as expected with red-detuned off-resonant light -- and the width $\Delta \omega$ of the distribution increases. At low trap depth $U_0 = 310\pm 10\,\mu$K the distribution has mainly one lobe, corresponding to the $\vert F=2\rangle \rightarrow \vert F'=3\rangle$ transition. For larger trap depths, two-photon couplings at 1559nm mix the excited state hyperfine levels and two extra lobes appear in the distributions, at lower frequencies, corresponding roughly to transitions $\vert F=2\rangle \rightarrow \vert F'=2\rangle$ and $\vert F=2\rangle \rightarrow \vert F'=1\rangle$. This illustrates the tunability of the inhomogeneous distribution with the intensity of the trapping field.
}
\label{fig:atomic_distribution}
\end{figure}

\begin{figure}[h]
\centering
\includegraphics[width=\columnwidth]{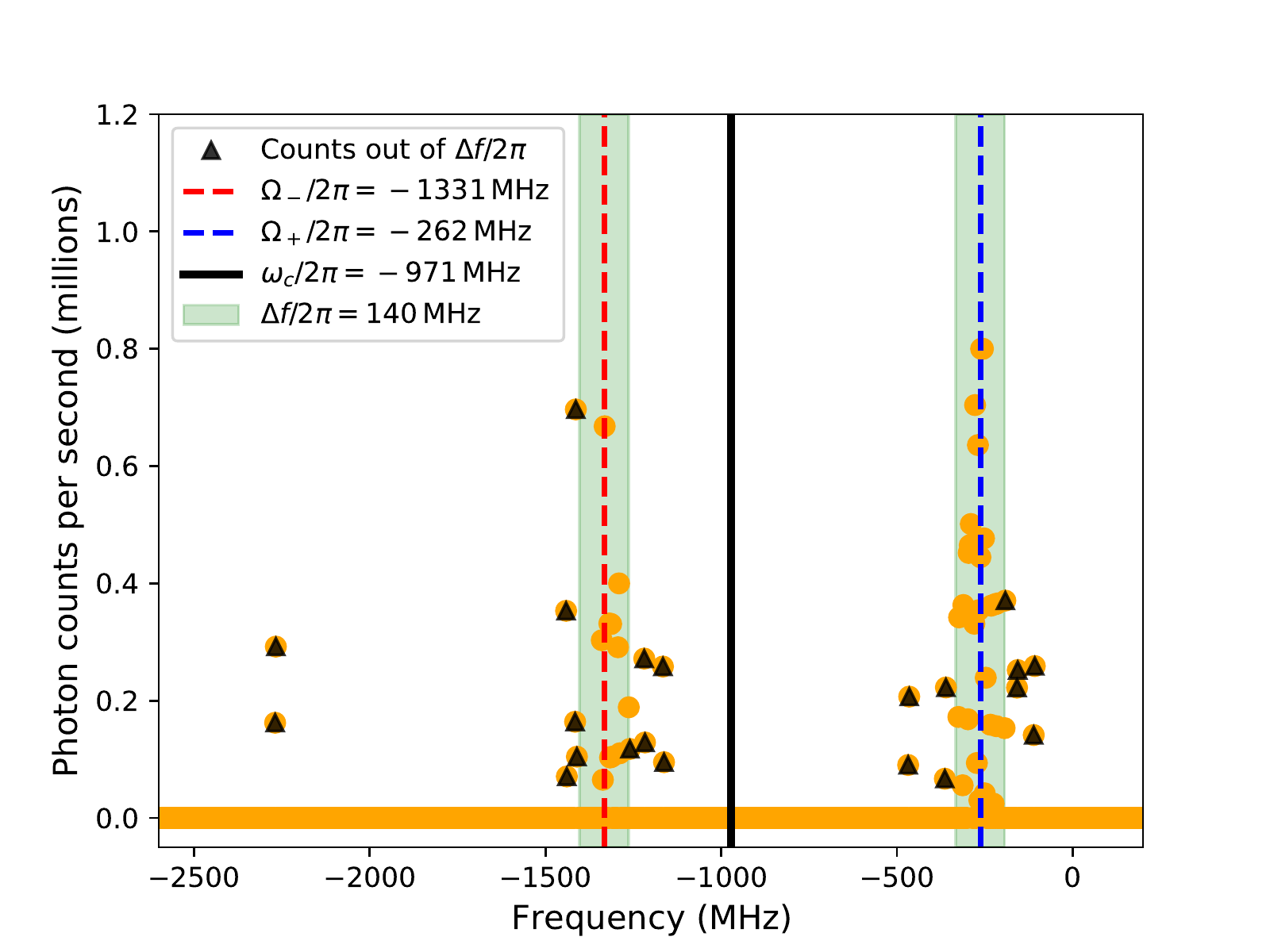}
\caption{\textbf{Single shot experimental spectrum.} As we probe the coupled system in the low excitation regime, we collect few photons in transmission and the spectrum is discretized (orange dots). For each spectrum, we compute the fraction $F_{out}$ of photons (identified with black triangles) that lies outside of a frequency window $\Delta f/2 \pi$ (green colored area), centered on each peak distribution (red and blue dashed lines).}
\label{fig:single_shot}
\end{figure}

\begin{figure}[h]
\centering
\includegraphics[width=\columnwidth]{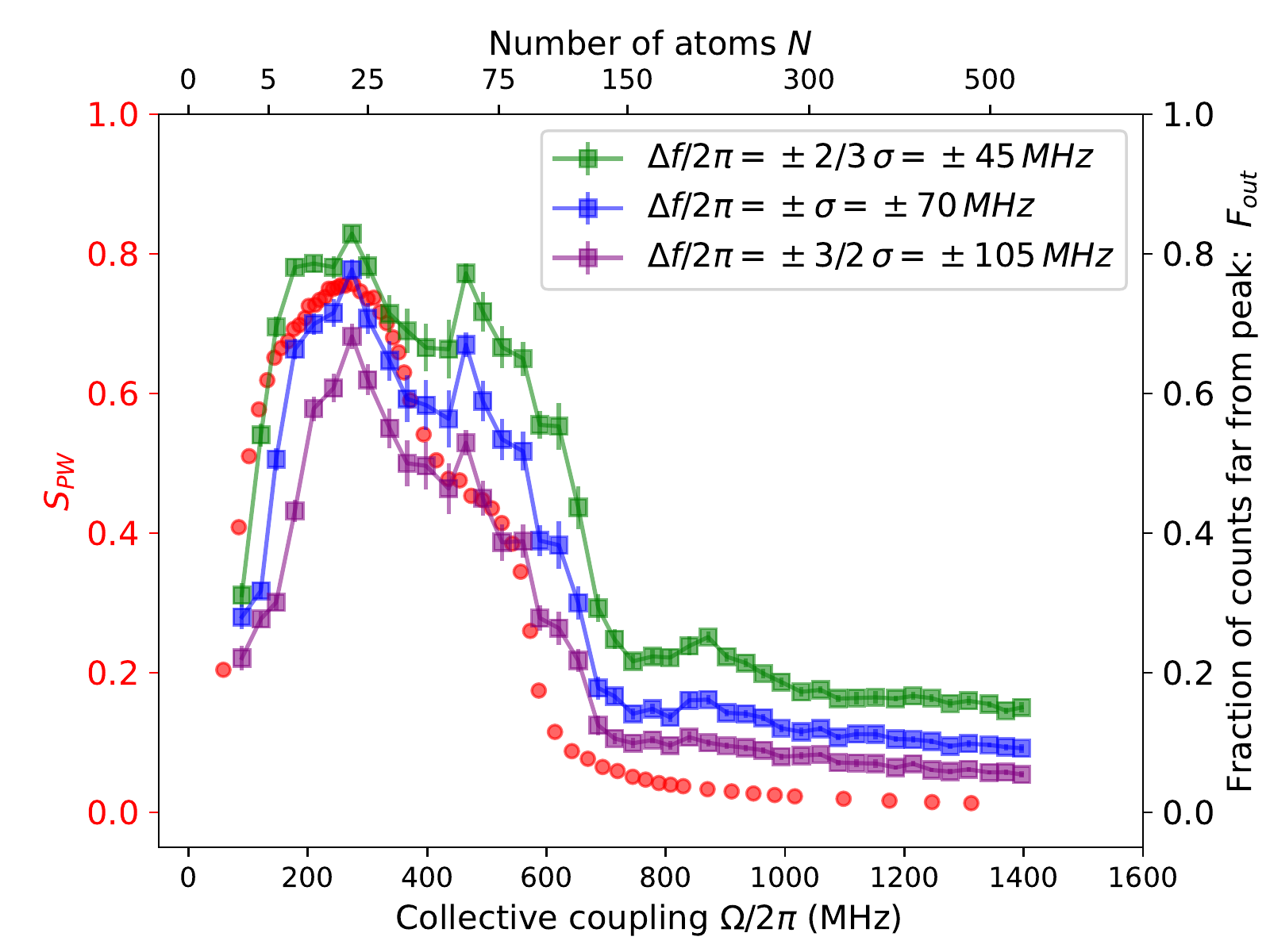}
\caption{\textbf{Robustness of $\mathbf{F_{out}}$ with respect to the size of the exclusion window.} Here we show $F_{out}$ for several values of $\Delta f / 2\pi$ chosen to define the exclusion window, together with the simulated sum of the two highest photonic weight, $S_{PW}$, as in Fig. 3 of the main text (red dots). The result is rather robust: when $\delta f / 2\pi$ decreases, the shape of $F_{out}$ remains the same, and is shifted upwards as expected.}
\label{fig:comparing_exclusion_window}
\end{figure}

\FloatBarrier

\pagebreak

\newpage
\onecolumngrid 

\begin{center}
    \vspace{0mm}
    %\hrule \vspace{5mm}
    \large{\textbf{SUPPLEMENTARY INFORMATION}}
    \vspace{3mm}
\end{center}

\twocolumngrid 

\setcounter{figure}{0}

\renewcommand{\thefigure}{S\arabic{figure}}

\section{Effect of the 1559$\,$nm light on the $5P_{3/2}$ manifold}

Following \cite{Arora2007, LeKien2013}, we describe the combined effects of lightshifts and level mixing within the $5P_{3/2}$ manifold induced by the 1559$\,$nm light by a Stark operator $\hat{V}$, whose matrix elements in the 
$ \vert F,m_F \rangle $ basis are given by:
\begin{equation}
\langle F',m_F' \vert \hat{V} \vert F,m_F\rangle= \sum_{\vert k\rangle } \frac{\langle F',m_F'\vert \hat{\mathbf{d}} \cdot \mathbf E \vert k\rangle \langle k \vert \hat{\mathbf{d}}
 \cdot \mathbf E \vert F,m_F\rangle}{-\hbar\Delta_{k}} \label{eq_V}
\end{equation}
where $\Delta_{k}$ is defined as:
$$
\frac{1}{\Delta_{k}}=\frac{1}{(\omega_{k}-\omega_{5P_{3/2}})-\omega}+\frac{1}{(\omega_{k}-\omega_{5P_{3/2}})+\omega} \;,
$$
all relevant states $\vert k\rangle$ being much farther in energy than the hyperfine splitting. In our case, given the value of the reduced matrix element and the detuning from the 1559$\,$nnm light of each atomic line \cite{Arora2007}, we expect most ($>99\%$) of the contribution in the sum of equation (\ref{eq_V})  to result from $4D_{5/2}$, $4D_{3/2}$, $6S_{1/2}$ and $5S_{1/2}$, so we keep only these lines in the simulations.

To compute the matrix elements $\langle F',m_F'\vert \hat{\mathbf{d}}  \cdot \mathbf E \vert F,m_F\rangle$ for a given atomic line and quantization axis $\mathbf{e}_z$, we express the electric field of the 1559$\,$nm trapping light $\mathbf E =(E_x,E_y,E_z)$ in the spherical basis:
\begin{equation}
\mathbf E = \sum_{q=-1,0,1} E_q \mathbf{e}_q \;, \label{eq_spherical}
\end{equation}
where the $\mathbf{e}_q$ are unitary vectors and the $E_q$ are the ``pi'' and ``sigma'' components of $\mathbf E$ related to $E_{x,y,z}$ by:
\begin{equation*}
E_0=E_z \quad \textrm{and} \quad E_{\pm 1}=(\pm E_x+iE_y)/\sqrt{2} \;.
\end{equation*}
The matrix elements of $ \hat{d}_q=\hat{\mathbf{d}} \cdot \mathbf{e}_q$ are then computed using the Wigner-Eckart theorem, and reduced matrix elements taken from reference \cite{Arora2007}. In our typical experimental situation, the magnetic field is along the cavity axis $z$, and the polarization of the trapping light is linear along the $x$ direction, corresponding to $E_0/|\mathbf E|=0$, $E_1/\vert\mathbf E \vert=1/\sqrt{2}$ and $E_{-1}/\vert \mathbf E\vert =-1/\sqrt{2}$.

\section{Cavity Transmission Spectrum}

\subsection{Inhomogeneous Tavis-Cummings model}

The coherent evolution of the coupled atom-cavity system is described by a multilevel version of the standard Tavis-Cummings Hamiltonian (with $\hbar \equiv 1$):
\begin{equation} \label{eq:tavis_cummings}
H=H_\textrm{cav}+H_\textrm{at}+H_\textrm{int},
\end{equation}
where:
\begin{itemize}
\item $H_\textrm{cav}=\omega_c a^\dag a$ is the Hamiltonian of the cavity mode;
\item $H_\textrm{at}= \sum_{k}  \sum_{j}  \omega_{k,j} \sigma_{k,j}^{+} \sigma_{k,j}^{-}$ is the Hamiltonian of the multilevel atomic ensemble dressed by the 1559$\,$nm light; the operator $\sigma_{k,j}^{+}$ is by definition equal to $\vert \psi_{k,j}\rangle \langle k:2,2 \vert$ and $\sigma_{k,j}^{-}=\left( \sigma_{k,j}^{+} \right)^\dag$;
\item $H_\textrm{int} = \sum_{k} \hbar g_{(k)} \left( \sigma_{k}^{-} a^\dag+\sigma_{k}^{+} a \right)$
describes the interaction between the atoms and the cavity mode in the rotating wave approximation, where:

$g_{k}=g(x_k,y_k,z_k)$ is the coupling constant of the atom $k$, at the position $(x_k,y_k,z_k)$, and $\sigma_{k}^{+}=\vert k:3,3\rangle \langle k:2,2 \vert$.
We can express $H_\textrm{int}$ in the following form :

$\displaystyle H_\textrm{int}= \sum_{k} \sum_{j}  \left( g_{k,j}  \sigma_{k,j}^{+} a  + g_{k,j} \sigma_{k,j}^{-} a^\dag \right)$ 

with $g_{k,j}=g_{k} \langle \psi_{k,j}  \vert k:3,3\rangle$.
\end{itemize}

In the low excitation limit, where the average photon number inside the cavity is much smaller than unity, the Hilbert space can be reduced to the subspace with one excitation: $\left\{ \vert 1, G \rangle, \vert 0, W\rangle \right\}$. $\vert 1, G \rangle$  corresponds to the state with one photon in the cavity mode and all atoms in the ground state $\vert G\rangle=\vert 1:2,2,\ldots,N:2,2\rangle$. $\vert 0, W\rangle$ is the state with zero photon in the cavity mode and one atomic excitation $\vert W\rangle = \Sigma^{\dagger} \vert G\rangle$, where $\Sigma^{\dagger}$ is the collective excitation operator, given by  $\Sigma^{\dagger} = \sum_{k} \frac{g_k}{\Omega} \sigma_{k}^{+}$.

\subsection{Trap and cavity parameters}

Deep into the cavity-protected regime, we observe two transmission peaks separated by twice the collective coupling $\Omega$. By also measuring the number $N$ of atoms in the same sequence using absorption imaging, we have access to the experimental single-atom effective coupling strength $g_{exp} = \frac{\Omega}{\sqrt{N}}$. In the range of depths and temperatures used in the experiment, we measure coupling strength $g_{exp}/2\pi$ between $55$ and $68\,$MHz (see Table I). The computed average value $g_{sim}$ of the simulated frequency distribution is in good agreement with the measured data, using the calibration procedure described in the following section for the trap depth $U_0$, and time-of-flight measurements of the temperature $T$.

\begin{table}%[H] add [H] placement to break table across pages
\caption{Trap and cavity parameters}
\label{trap_parameters}

\begin{ruledtabular}
\centering
\begin{tabular}{|c|c|c|c|}
$U$ ($\mu$K) & $T$ ($\mu$K)  & $g_{sim}/2\pi$ (MHz)   & $g_{exp}/2\pi$ (MHz) \\
\hline
310$\pm$10  & 50$\pm$3  & 57 & 56$\pm$4  \\
\hline
710$\pm$20 & 100$\pm$10 & 60  & 62$\pm$4  \\
\hline
1040$\pm$30  & 140$\pm$20  & 60   & 66$\pm$4  \\
\hline
1400$\pm$30  & 190$\pm$20  & 60  & 67$\pm$4  
\end{tabular}
\end{ruledtabular}

\end{table}

\subsection{Cavity detuning and light-shift calibration}
\label{Lsmeasurement}

In the simple case of a two-level atom with frequency $\omega_A$ strongly coupled to a cavity with frequency $\omega_c$, the two polaritons are centered on the frequency $(\omega_A+\omega_c)/2$ \cite{Haroche2006}. In particular, when the cavity is tuned to resonance with the atomic frequency ($\omega_c=\omega_A$), the center of the transmission doublet coincides with the position of the empty cavity. Our simulations for an inhomogeneous atomic ensemble reveal a similar (and more general) condition: the transmission doublet is centered on the empty cavity when its frequency is equal to the average (lightshifted) atomic frequency, defined by $\displaystyle \bar{\omega}_A=\sum_{k,j} \left(g_{k,j}\right)^2 \omega_{k,j}/\Omega^2$. To verify this, we simulate, for various atom-cavity detunings,  transmission spectra, for N=500 atoms, deep in the cavity-protected regime. From each spectrum we extract the central frequency of the Rabi doublet, and average it over 40 realizations. We then fit the average central frequency against detuning with a linear function, which gives us the detuning for which the cavity frequency is at the center of the Rabi doublet. This atom-cavity detuning is compatible with the average atomic lightshift, computed over 20.000 atoms.

Based on this observation, we experimentally tune the cavity mode frequency to resonance with the average atomic frequency by centering the empty cavity on the transmission doublet in the regime of strong collective coupling. Interestingly, knowing the frequency of the tuned cavity mode provides a precise measurement of the average lightshift, and also of the trap depth based on independent temperature measurements (in practice, we calculate the average lightshift for various trap depths until it coincides with the experimental value up to 1$\%$, which is below the experimental uncertainties). Experimental averaged frequencies $\bar{\omega}_{A, exp}$ are shown in Table II.  
The values of $U_0$ are compatible with the measurements performed by direct intensity calibration within the uncertainties margins.

\begin{table}[h]% add [H] placement to break table across pages
\caption{Calibration of trap depth $U_0$}\label{averaged_frequency}
\begin{ruledtabular}
\begin{center} 
\begin{tabular}{|c|c|c|}
$\bar{\omega}_{A}$ (experimental) & $T$ (experimental) & $U_{0}$ (simulation) \\
\hline
-$280 \pm 10\,$MHz & $50\pm 3\,\mu$K  & $310 \pm 10\,\mu$K  \\
\hline
-$660\pm 10\,$MHZ & $100\pm 10\,\mu$K   & $710 \pm 20\,\mu$K  \\
\hline
-$970\pm 20\,$MHz & $140\pm 20\,\mu$K  & $1040\pm 30\,\mu$K  \\
\hline
-$1300\pm 40\,$ MHz & $190\pm 20\,\mu$K  & $1400\pm 30\,\mu$K   
\end{tabular}
\end{center} 
\end{ruledtabular}

\end{table}

\pagebreak

\newpage

\onecolumngrid 

\begin{widetext}

\section{Polaritons Modulation}

\subsection{Hamiltonian of the modulated atom-cavity system}

The data for the polariton modulation have been acquired in the fully protected regime ($\Omega \gg \Delta \omega$). This allows us to consider a simplified model where all atoms have the same atomic frequency $\omega_A(t)$ dressed by the time-dependent 1559$\,$nm dipole light. For this homogeneous system and considering only the low excitation limit, we can restrict the dynamics of the coupled cavity-atoms system to the symmetric subspace of the first excitation manifold spanned by the basis states  $\left\{ \vert 1, G \rangle, |0, W\rangle \right\}$ with $\vert G\rangle=|g_{1}...g_{N}\rangle$ and $ \vert W\rangle = \Sigma^{\dagger} \vert G\rangle$, where $\Sigma^{\dagger}$ is the collective excitation operator, given by  $\Sigma^{\dagger} = \sum_{k=1}^N \frac{g_{k}}{\Omega} \sigma_{k}^{+}$, with $\Omega$ the collective atom-cavity coupling. In this configuration, the $N-1$ dark states do not play a role in the time evolution of the system. The Hamiltonian of the system is then given by the time-dependent version of the standard Tavis-Cummings Hamiltonian

\begin{equation} \label{eq:tavis_cummings_temp}
H(t)=  \omega_c a^{\dagger} a + \omega_A(t) \Sigma^{\dagger} \Sigma +  \Omega \left( \Sigma a^{\dagger} + \Sigma^{\dagger} a \right).
\end{equation}

By coupling two frequency components of the trapping light within the cavity bandwidth, we produce a beating that induces a modulation of the trapping dipole light at a frequency $\omega_m /2 \pi$:  $I_D(t)=I_D^0+ \delta I \cos(\omega_m t + \phi)$. This leads to a modulation of the trapping potential $U(t)=U^0+ \delta U \cos(\omega_m t + \phi)$. As this modulation is quite fast in the 100$\,$MHz range, the positions of the atoms in the trap cannot respond to the modulation and we can consider the $g$ coupling distribution is constant. However, the light-shifted atomic frequency is instantaneously modified, and  we obtain at first order a modulation  $\omega_A(t) = \omega_A^0 + \delta \omega_A(t)$ with $  \delta \omega_A(t) = \beta_o \omega_m \cos(\omega_m t + \phi)$, where we define $\beta_o$ as the "optical" modulation index.  The Hamiltonian can then be written as 
\begin{equation} \label{eq:tavis_cummings_temp2}
H(t)=  \omega_c\, a^{\dagger}a + \omega_A^0\, \Sigma^{\dagger} \Sigma +  \Omega\, \left( \Sigma a^{\dagger} + \Sigma^{\dagger} a \right) + \delta \omega_A(t)\, \Sigma^{\dagger} \Sigma.
\end{equation}

By moving into  the rotating frame of the probe laser with frequency $\omega_p$, we can express the Hamiltonian as
\begin{eqnarray} \label{eq:tavis_cummings_temp3}
\tilde{H}(t) &=&  \Delta\omega_c\, a^{\dagger} a \,+ \, \Delta\omega_A^0\, \Sigma^{\dagger} \Sigma \,+\,  \Omega\,\left( \Sigma a^{\dagger} + \Sigma^{\dagger} a \right)
 +  \delta \omega_A(t)\, \Sigma^\dag \Sigma,
\end{eqnarray}
with $\Delta \omega_c=\omega_c-\omega_p$ and $ \Delta \omega_A^0=\omega_A^0-\omega_p$.
\vspace{0.25 cm}

The Hamiltonian is divided into the standard Tavis-Cummings Hamiltonian and a time-dependent perturbation:  $\tilde{H}(t)=\tilde{H}_{TS}+\tilde{V}(t)$. The eigenvalues of $\tilde{H}_{TS}$ are given by 
\begin{equation}
E_{\pm}= \frac{\omega_A^0+\omega_c}{2}-\omega_p \pm \sqrt{ \Omega^2 + \left( \frac{\omega_A^0-\omega_c} {2} \right)^2}.
\end{equation}

In the following, we consider only the resonant case $\omega_c = \omega_A^0 = \omega_0$. As the frequency modulation $\delta \omega_A(t)$ is small compared to  $\Omega$, we express $\tilde{H}(t)$ in the polariton basis $\lbrace \vert P_+ \rangle,\vert P_- \rangle \rbrace$ of the unmodulated system defined by
\begin{equation}
\vert P_\pm \rangle=\frac{1}{\sqrt{2}} \left(\vert 1 , G\rangle \pm \vert 0, W \rangle \right).
\end{equation}

We obtain
\begin{align}
\tilde{H}(t)= 
\begin{pmatrix*}
\omega_0-\omega_p + \Omega & 0\\
0 & \omega_0-\omega_p - \Omega \\
\end{pmatrix*}
+\frac{\delta \omega_A(t)}{2}
\begin{pmatrix*}[l]
1 & -1\\
-1 & 1
\end{pmatrix*}
\label{Equation_hamiltoniendans+-}
\end{align}

We see in this expression that the modulation of the emitters frequency modifies the eigenfrequencies of the Tavis-cummings Hamiltonian and also induces a coupling between the two polaritonic states $\lbrace \vert B_+ \rangle,\vert B_- \rangle \rbrace$, which are not any more the eigenstates of the system.

\subsection{Analytical expressions}

To underline the main physical effect of the modulation, we first consider the situation where initially one excitation is stored in the emitter-cavity system and we calculate the Rabi oscillation and cavity spectrum, which is given by the following relation \cite{Gardiner2004}
\begin{align}
S(\omega) = \frac{1}{2 \pi} \int_{0}^{\infty} dt_1 \int_{0}^{\infty} dt_2  e^{i \omega (t_2-t_1)} \,  \langle a(t_1) a^{\dagger}(t_2) \rangle  
 \propto \int_{0}^{\infty} dt_1 \int_{0}^{\infty} dt_2  e^{i \omega (t_2-t_1)} \,  \langle  P^{\dagger}(t_1) P(t_2) \rangle
\end{align}

Following the approach of \cite{Law1995}, we consider the system in a pure state 
\begin{equation}
\vert \psi (t)\rangle = c_g(t) \vert 1 , G\rangle + c_e(t)\vert 0, W \rangle 
\end{equation}
and calculate its temporal evolution. 
The damping of the cavity photons and losses by spontaneous emission are taken into account by using an "effective" non-hermitian Hamilonian $\tilde{H}_{eff}(t)$ obtained by replacing $\omega_A\rightarrow \omega_A - i \gamma$ and $\omega_c \rightarrow \omega_c - i \kappa$ \cite{Diniz2011b,Kurucz2011a} .

In the rotating frame, we can write in the polariton basis :
\begin{equation}
\vert \tilde{\psi} (t)\rangle = \tilde{c}_+(t)  \vert P_+ \rangle  +  \tilde{c}_-(t) \vert P_- \rangle
\end{equation}
with $\tilde{c}_\pm(t) =(\tilde{c}_g(t) \pm \tilde{c}_e(t))/\sqrt(2) $
and $\tilde{H}_{eff}(t)$ is given by
\begin{align}
\tilde{H}_{eff}(t)=
\begin{pmatrix*}[c]
\omega_0-\omega_p + \Omega - i \gamma_s & \,\,\,\,- i \gamma_d\\
- i \gamma_d  &  \omega_0-\omega_p-\Omega -i \gamma_s \\
\end{pmatrix*} 
+ \frac{\delta \omega_A(t)}{2}
\begin{pmatrix*}[c]
1&-1\\
-1& 1
\end{pmatrix*}
\label{Equation_hamiltoniendans_losses}
\end{align}
with $\gamma_s=\frac{\kappa+\gamma}{2}$  et $\gamma_d = \frac{\kappa - \gamma}{2}$.
\vspace{0.25 cm}

We now  introduce the two functions $\tilde{b}_+(t)$ and $\tilde{b}_-(t)$ defined by
\begin{eqnarray}
\vert \tilde{\psi} (t)\rangle = \tilde{b}_+(t) e^{-\gamma_s t}  e^{-i(\omega_0-\omega_p)t} e^{-i\frac{\beta_o}{2} \sin \left(\omega_m t + \phi \right)} e^{-i \Omega t}  \vert B_+ \rangle 
+  \tilde{b}_-(t)  e^{-\gamma_s t} e^{-i(\omega_0-\omega_p)t}  e^{-i\frac{\beta_o}{2} \sin \left(\omega_m t + \phi \right)}  e^{i \Omega t} \vert B_- \rangle
\end{eqnarray}
which are solutions of the two differential equations:
 
\begin{equation}
\renewcommand{\arraystretch}{2} % Default value: 1
\left\lbrace
\begin{array}{cc}
i \frac{\partial}{\partial t} \tilde{b}_+ =& - \left( \frac{\beta_o \omega_m }{2} \cos \left( \omega_m t +\phi \right) + i \gamma_d \right) e^{2i\Omega t} \, \tilde{b}_-  \\ 
i \frac{\partial}{\partial t} \tilde{b}_- =& - \left( \frac{\beta_o \omega_m }{2} \cos \left( \omega_m t +\phi \right) + i \gamma_d \right)  e^{-2 i \Omega t} \, \tilde{b}_+
\end{array}\label{equa_diff}
\right.
\end{equation}

\subsubsection{Non-resonant case $\omega_m \ll \Omega$ and small modulation $\beta_o \ll 1$}

To calculate the cavity spectrum, we consider first the situation where $\Omega \gg \beta_o \omega_m, \gamma_d$. This is a good approximation of the experiment, which is in a clearly distinct regime than the experiment of Ref. \cite{Clark2019}. 
We can then neglect the fast rotating terms  $e^{\pm 2i\Omega t} b_{\pm}$ in the equations \ref{equa_diff}, which is equivalent to neglect the coupling between the two polaritons. In this case, the solution of the above equations are straightforward:
\begin{equation}
\tilde{b}_{\pm}(t)= \tilde{b}_{\pm}(0) \, \Rightarrow  
\tilde{c}_{\pm}(t)= e^{\mp i \Omega t} e^{-\gamma_s t} e^{-i(\omega_0-\omega_p)t} e^{-i \frac{\beta_o}{2} \sin \left(\omega_m t +\phi \right)} e^{i \frac{\beta_o}{2} \sin \left(\phi \right)} \,\, \tilde{c}_{\pm}(0)
\end{equation}

Going back to the lab frame, starting from the initial state $\vert \psi (0)\rangle = \vert 1 , G\rangle $, we compute the coefficient of the states $\vert 1 , G\rangle$ and  $\vert 0, W \rangle $
\begin{equation}
\renewcommand{\arraystretch}{2} % Default value: 1
\left\lbrace
\begin{array}{cc}
c_g(t)&=  e^{i \frac{\beta_o}{2} \sin \left(\phi \right)} e^{-i\omega_0 t} e^{-\gamma_s t} e^{-i\frac{\beta_o}{2} \sin \left(\omega_mt +\phi \right)} \cos (\Omega t ) \\
c_e(t)&= -i e^{i \frac{\beta_o}{2} \sin \left(\phi \right)} e^{-i\omega_0 t} e^{-\gamma_s t}  e^{-i\frac{\beta_o}{2} \sin \left(\omega_mt +\phi \right)} \sin (\Omega t ) 
\end{array}
\right.
\end{equation}\label{solution_non_resonant}
We obtain Rabi oscillations at a frequency $2 \Omega$ between these two states as in the unmodulated case. 

However, the cavity spectrum is modified due to the modulation of the eigenenergies.  Noting  that $\langle  P^{\dagger}(t_1) P(t_2) \rangle$= $c_e^*(t_1) c_e(t_2)$, we can indeed write the cavity spectrum as
\begin{equation}
S(\omega) \propto \int_{0}^{\infty} dt_1 \int_{0}^{\infty} dt_2  e^{i\omega (t_2-t_1)} \, c_e^*(t_1) c_e(t_2) 
\propto  \int_{0}^{\infty} dt_1 c_e^*(t_1) e^{-i\omega t_1} \int_{0}^{\infty} dt_2  c_e(t_2) e^{i\omega t_2} 
\end{equation}

The integrals can be expanded using $\displaystyle e^{-i \beta \sin \left(\omega_m t +\phi \right)} = \sum_{n=-\infty}^{+\infty} J_n \left(\beta \right)  e^{- i n \omega_m t} e^{- i n \phi}  $ and after integration, the cavity spectrum is given by:

\begin{equation}\label{cavity_spectrum_non_resonant}
S(\omega)  \propto  \left \vert \sum_{n=-\infty}^{+\infty}  J_n  \left(\beta_o/2 \right)  e^{- i n \phi} 
 \left( \frac{1}{\omega-\omega_0-n\omega_m+ \Omega +i \gamma_s} 
 - \frac{1}{\omega-\omega_0-n\omega_m - \Omega+i \gamma_s} \right) \right \vert^2 .
\end{equation}

We see on this expression that multiple frequencies are present in the spectrum, given by $\omega=\pm \Omega + n \omega_m$ with $n$ integer. The peaks of the doublet centered on $\pm \Omega$ acquires multiple sidebands spaced by the modulation frequency $\omega_m$.

\subsubsection{Near-resonant case $\omega_m \simeq 2 \Omega$}

An interesting case occurs when the modulation frequency is close to twice the collective coupling $\omega_m \simeq 2 \Omega $ with $\Omega \gg \gamma_d$. In the equations \ref{equa_diff}, we can then only keep the slowly varying term (rotating wave approximation) and obtain :
\begin{equation}
\renewcommand{\arraystretch}{2} % Default value: 1
\left\lbrace
\begin{array}{cc}
i \frac{\partial}{\partial t} \tilde{b}_+ &= - 
\frac{\beta_o \omega_m }{4} e^{i \left( (2 \Omega-\omega_m) t -\phi \right)} \, \tilde{b}_-  \\
i \frac{\partial}{\partial t} \tilde{b}_- &= - \frac{\beta_o \omega_m }{4} e^{-i \left( (2 \Omega-\omega_m) t -\phi \right)} \, \tilde{b}_+ .
\end{array}
\right.
\end{equation}\label{equad_diff_res}

After integration, we obtain the population of the polaritonic states (starting from the initial state $\vert \psi (0)\rangle = \vert 1 , G\rangle $):
\begin{equation}
\renewcommand{\arraystretch}{2} % Default value: 1
\left\lbrace
\begin{array}{cc}
\vert c_+(t)\vert ^2 &= \frac{1}{2} \left(1 - \sin(\phi) \sin\left( \beta_o \omega_m t/2 \right)\right) e^{- 2 \gamma_s t} \\ 
\vert c_-(t)\vert ^2 &= \frac{1}{2} \left(1 + \sin(\phi) \sin\left( \beta_o \omega_m t/2 \right)\right) e^{- 2 \gamma_s t}.
\end{array}
\right.
\end{equation}\label{equad_diff_sol_res_e_g}
We clearly see in these expressions that a very effective polaritonic populations transfert occurs  at the frequency $\Omega_m=\beta_o \omega_m/2$ induced by the modulation of the atomic frequency (see Fig. \ref{Rabi_oscillations_resonant}). This provides a way to manipulate the polaritons on very fast time scales. We note also that the oscillations depend on the phase of the modulation and are maximal for $\phi=\pi/2$ \cite{Law1995}.

\begin{figure}
\centering
\includegraphics[width=\columnwidth ]{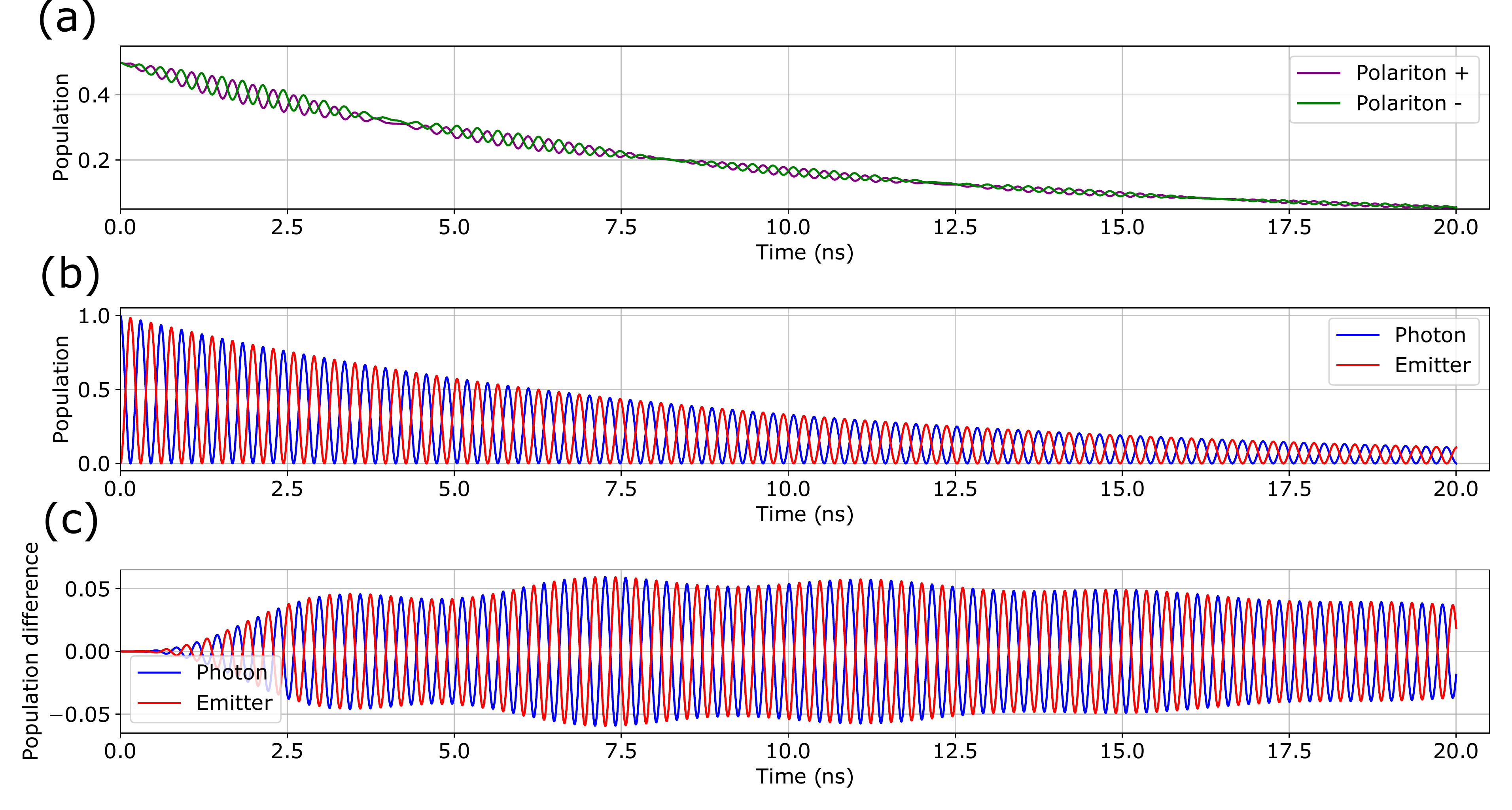}
\caption{ \textbf{Simulation results in the case of a resonant modulation : $\mathbf{\omega_m = 2 \Omega}$.} The parameters are $\Omega /2 \pi= 1634\,$MHz, $\kappa/2 \pi= 14.6\,$MHz,  $\gamma/2 \pi= 3\,$MHz, $\beta_o= 0.17$, $\phi=\pi/2$. (a) Populations of the polaritonic states $\vert B_+\rangle$ $\vert B_-\rangle$ defined in the steady state case. The modulation induces a coupling between the two polaritons leading to oscillations at the frequency $\beta_o \omega_m  /2 $. (b) Modulated Rabi oscillations of the populations of $\vert 1, G \rangle$ and $\vert 0, W\rangle$  (c) Populations difference between the modulated and unmodulated case. (d) Cavity Spectrum for a resonant modulation  $\omega_m=2 \Omega$ ,$\Omega/2 \pi= 1634\,$MHz, $\kappa/2 \pi= 14.6\,$MHz,  $\gamma/2 \pi= 3\,$MHz, $\beta_o= 0.17$, $\phi=\ pi/2$.}
\label{Rabi_oscillations_resonant}
\end{figure}
\vspace{0.5 cm}

The expressions of the coefficient $c_g(t)$ and $c_e(t)$ for a phase $\phi=\pi/2$ are given by:
\begin{equation}
\renewcommand{\arraystretch}{2} % Default value: 1
\left\lbrace
\begin{array}{cc} 
\vert c_g(t)\vert ^2&=  \left( \cos^2(\Omega_m t/2) \cos^2(\Omega t)+\sin^2(\Omega_m t/2) \sin^2(\Omega t) \right)e^{- 2 \gamma_s t}\\
\vert c_e(t)\vert ^2&= \left( \sin^2(\Omega_m t/2) \cos^2(\Omega t)+\cos^2(\Omega_m t/2) \sin^2(\Omega t) \right)e^{- 2 \gamma_s t}. 
\end{array}
\right.
\end{equation}\label{Rabi_res_sol}
The populations of these two states feature an oscillation at the frequency $2 \Omega$ with a modulation at the frequency $\Omega_m = \beta_o \omega_m/2$ and a damping at rate $2 \gamma_s=\kappa+\gamma$ (see Fig.\ref{Rabi_oscillations_resonant}).\\

We calculate the spectrum in this regime, which gives for a resonant modulation $\omega_m=2\Omega$:
\begin{align}
S(\omega) &\propto  \left\vert \sum_{n=-\infty}^{+\infty}  J_n  \left(\beta_o/2 \right)  e^{- i n \phi} \left( \frac{1- e^{-i \phi}}{\omega-\omega_0-2 n \Omega+ + \Omega( 1+\beta_o)  +i \gamma_s} +
 \frac{1+e^{-i \phi}}{\omega-\omega_0-2 n \Omega + \Omega( 1-\beta_o) + i \gamma_s}  \right.  \right.  \\ \nonumber
&- \left. \left. \frac{1+ e^{i \phi}}{\omega-\omega_0-2 n \Omega - \Omega( 1+\beta_o) +i \gamma_s} -
 \frac{1- e^{i \phi}}{\omega-\omega_0 - 2 n \Omega  - \Omega( 1-\beta_o) +i \gamma_s} \right) \right \vert^2.
\end{align}

For small modulation depth, the spectrum is dominated by four frequencies $\Omega( 1 \pm \beta_o)$ and $-\Omega( 1 \pm \beta_o)$, this is the situation depicted on Fig.Fig.\ref{Rabi_oscillations_resonant}.d. The different height of the peaks results from the interference between the sidebands of the different terms contributing to the spectrum.

\subsubsection{Non-resonant case $\omega_m < 2 \Omega$ with non-negligible modulation  $\beta_o \geq 1$}

In this regime, which corresponds to our experiment, we cannot completely neglect the coupling between the polaritons, so it is not legitimate to perform the rotating wave approximation as it has been done previously. We have integrated numerically the master equation for an initial state with one photon stored in the cavity mode (see Fig. \ref{Rabi_oscillations}).
We first see that oscillations occur between the two polaritons at a frequency of about $2 \Omega$. The amplitude of these oscillations is rather small (roughly 4\%) due to the fact that the modulation frequency is far-detuned from resonance. A modulation of these oscillations at a frequency $2 \omega_m$ is also present, stemming from the anti-resonant term. We can see also Rabi oscillations between  the states $|1, G \rangle$ and $|0, W\rangle$  with a slight modulation at the frequency $2 \omega_m/2 \pi$, clearly visible on the population difference between the modulated and unmodulated case.

\begin{figure}
\centering
\includegraphics[width=0.9 \columnwidth]{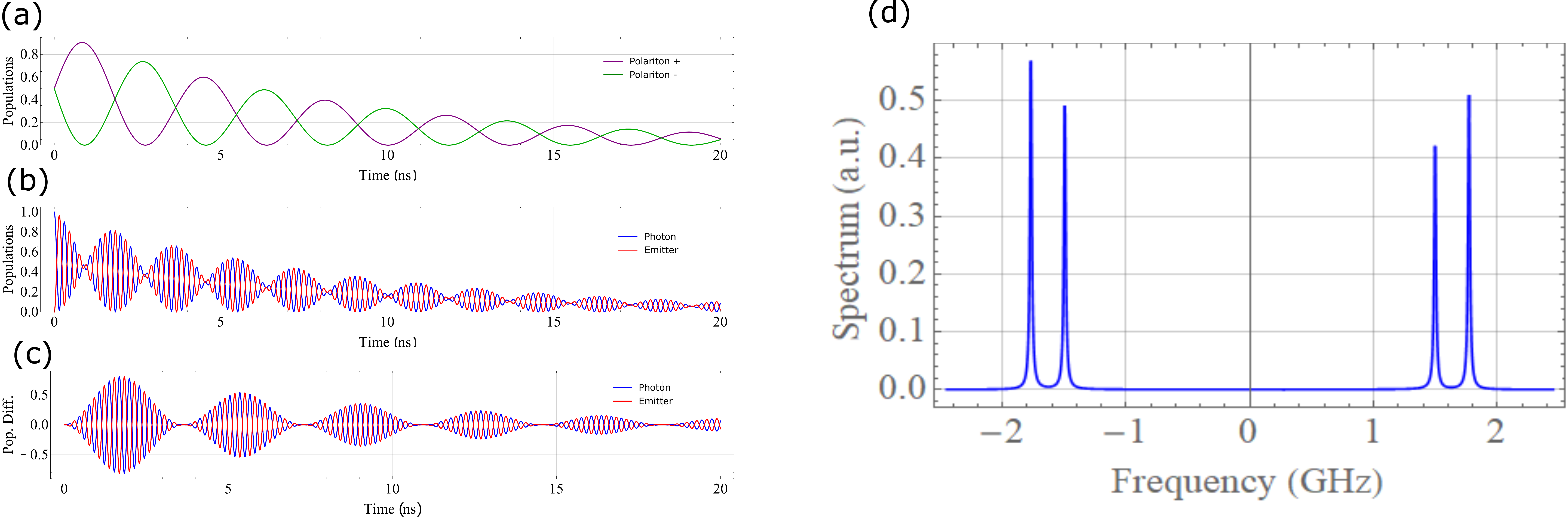}
\caption{\textbf{Simulation for our experimental parameters:} $\omega_m= 122 \,$MHz , $g/2 \pi= 1634\,$MHz, $\kappa/2 \pi= 14.6\,$MHz,  $\gamma/2 \pi= 3\,$MHz, $\beta_o= 2.17$, $\phi=\pi/2$.  (a) Populations of the polaritonic states $\vert B_+\rangle$ $\vert B_-\rangle$ defined in the steady state case. The modulation induces a coupling between the two polaritons leading to oscillations at the frequency $\beta_o \omega_m  /2$. (b) Modulated Rabi oscillations of the populations of $\vert 1, G \rangle$ and $\vert 0, W\rangle$  (c) Populations difference between the modulated and unmodulated case.}
\label{Rabi_oscillations}
\end{figure}

\subsubsection{Fitting function for experimental spectra}

For our experimental parameters, the coupling between the polaritons is relatively weak, so we derive an analytical formula to be able to fit experimental data efficiently.
We start from the expression \ref{cavity_spectrum_non_resonant} of the cavity spectrum. It can be developed as 
 \begin{eqnarray}
S(\omega)  \propto   \sum_n \sum_l  J_n  \left(\beta_o/2 \right) J_l  \left(\beta_o/2 \right) e^{- i (n-l) \phi} \nonumber
& &  \left( \frac{1}{(\omega-\omega_0-n\omega_m+ \Omega )+i \gamma_s} \nonumber 
 - \frac{1}{(\omega-\omega_0-n\omega_m - \Omega)+i \gamma_s} \right) \times \nonumber \\
 & &\left( \frac{1}{(\omega-\omega_0-l \omega_m+ \Omega )-i \gamma_s} 
 - \frac{1}{(\omega-\omega_0-l \omega_m - \Omega)-i \gamma_s} \right)
\end{eqnarray}\label{cavity_spectrum_exp}

In the experiment, we do not directly measure the cavity spectrum obtained as a Fourier-transform of the temporal oscillation of an initial single excitation. We rather observe the cavity transmission of a weak laser probe swept in frequency, so the phase $\phi$ is not fixed, and in the equation (23), only the diagonal terms will contribute:
\begin{eqnarray}
S(\omega)  \propto   \sum_n   J_n^2  \left(\beta_o/2 \right)   
 \left \vert \frac{1}{(\omega-\omega_0-n\omega_m+ \Omega )+i \gamma_s} 
- \frac{1}{(\omega-\omega_0-n\omega_m - \Omega)+i \gamma_s} \right \vert ^2 
\end{eqnarray}\label{cavity_spectrum_exper}

As we are in the regime where $\Omega \gg \omega_m $, we can neglect the crossing terms in the last expression, and we end up with:
\begin{eqnarray}
S(\omega)  \propto   \sum_n   
 \frac{ J_n^2   \left(\beta_o/2 \right)}{(\omega-\omega_0-n\omega_m+ \Omega )^2 + \gamma_s^2} 
+ \sum_n  \frac{ J_n^2   \left(\beta_o/2 \right)}{(\omega-\omega_0-n\omega_m - \Omega)^2+ \gamma_s^2} 
\end{eqnarray}\label{cavity_spectrum_exper_inter_2}

\end{widetext}

\end{document}